\documentclass{aastex61}
\usepackage{color}
\usepackage{longtable}
\usepackage{threeparttablex} 

\newcolumntype{L}[1]{>{\raggedright\arraybackslash}p{#1}}                  
\newcolumntype{C}[1]{>{\centering\arraybackslash}p{#1}}

\begin{document}

\title{\sc The role of environment in galaxy evolution in the SERVS Survey I: density maps and cluster candidates} 

\author{Nick Krefting}
\affiliation{Department of Physics and Astronomy, Tufts University, 574 Boston Avenue, Medford, MA 02155, USA}
\correspondingauthor{Nick Krefting}
\email{nicholas.krefting@tufts.edu}

\author{Anna Sajina}
\affiliation{Department of Physics and Astronomy, Tufts University, 574 Boston Avenue, Medford, MA 02155, USA}

\author{Mark Lacy}
\affiliation{National Radio Astronomy Observatory, 520 Edgemont Road, Charlottesville, VA 22903, USA}

\author{Kristina Nyland}
\affiliation{National Research Council, resident at the Naval Research Laboratory, Washington, DC 20375, USA}

\author{Duncan Farrah}
\affiliation{Department of Physics and Astronomy, University of Hawaii, 2505 Correa Road, Honolulu, HI 96822,USA}
\affiliation{Institute for Astronomy, 2680 Woodlawn Drive, University of Hawaii, Honolulu, HI 96822, USA}

\author{Behnam Darvish}
\affiliation{California Institute of Technology, Pasadena, CA 91125, USA}

\author{Steven Duivenvoorden}
\affiliation{Astronomy Centre, Department of Physics and Astronomy, University of Sussex, Brighton BN1 9QH, UK}

\author{Ken Duncan}
\affiliation{Leiden Observatory, Leiden University, PO Box 9513, NL-2300 RA Leiden, the Netherlands}

\author{Violeta Gonzalez-Perez}
\affiliation{Astrophysics Research Institute, Liverpool John Moores University, 146 Brownlow Hill, Liverpool L3 5RF, UK}
\affiliation{Institute of Cosmology \& Gravitation, University of Portsmouth, Dennis Sciama Building, Portsmouth, PO1 3FX, UK}

%\affiliation{University of Portsmouth, UK}

\author{Claudia del P. Lagos}
\affiliation{International Centre for Radio Astronomy Research (ICRAR), M468, University of Western Australia, 35 Stirling Hwy, Crawley, WA 6009, Australia.}
\affiliation{ARC Centre of Excellence for All Sky Astrophysics in 3 Dimensions (ASTRO 3D)}
\affiliation{Cosmic Dawn Center (DAWN), Copenhagen, Denmark 0000-0003-3631-7176}

\author{Seb Oliver}
\affiliation{Astronomy Centre, Department of Physics and Astronomy, University of Sussex, Brighton BN1 9QH, UK}

\author{Raphael Shirley}
\affiliation{Astronomy Centre, Department of Physics and Astronomy, University of Sussex, Brighton BN1 9QH, UK}
\affiliation{Instituto de Astrofísica de Canarias, E-38205 La Laguna, Tenerife, Spain}
\affiliation{Dpto. Astrofísica, Universidad de La Laguna, E-38206 La Laguna, Tenerife, Spain}

\author{Mattia Vaccari}
\affiliation{Department of Physics and Astronomy, University of the Western Cape, Private Bag X17, Bellville 7535, South Africa}
\affiliation{INAF - Istituto di Radioastronomia, via Gobetti 101, I-40129 Bologna, Italy}

\begin{abstract}
We use photometric redshifts derived from new $u$-band through 4.5$\mu$m {\sl Spitzer} IRAC photometry in the 4.8\,deg$^2$ of the XMM-LSS field to construct surface density maps in the redshift range 0.1-1.5. Our density maps show evidence for large-scale structure in the form of filaments spanning several tens of Mpc. Using these maps, we identify 339 overdensities that our simulated lightcone analysis suggests are likely associated with
dark matter haloes with masses, $M_{\rm halo}$, log($M_{\rm halo}/M_{\odot})>$13.7. From this list of overdensities we recover 43 of 70 known X-ray detected and spectroscopically confirmed clusters. The missing X-ray clusters are largely at lower redshifts and lower masses than our target log($M_{\rm halo}/M_{\odot})>$13.7. The bulk of the overdensities are compact, but a quarter show extended morphologies which include likely projection effects, clusters embedded in apparent filaments as well as at least one potential cluster merger (at $z\sim1.28$). The strongest overdensity in our  highest redshift slice  (at $z\sim1.5$) shows a compact red galaxy core potentially implying a massive evolved cluster. 
\end{abstract}

\keywords{galaxies: general ---
galaxies: evolution --- galaxies:formation --- galaxies: photometry --- galaxies: statistics}

\section{Introduction} 
\label{sec:intro}

Many studies over the last few decades have shown that the local density in which a galaxy resides affects its growth, quenching and morphology \citep[e.g.][]{Cochrane2018}. The physical mechanisms through which the environment plays a role include gas accretion, feedback, and galaxy interactions. At low redshift, this environmental dependence is well known, with `red and dead' elliptical galaxies dominating in denser environments, while star-forming spirals are more commonly found in the field  \citep[e.g.][]{Dressler1980,Norberg2002,Peng2010}.  Higher-$z$ studies also show environmental trends such as the faster build-up and quenching of more massive galaxies in denser environments \citep[e.g.][]{vanderBurg2013,EtheringtonEtAl2017} and more generally the dependence of the specific star formation rate (SFR) on local environment \citep[e.g.][]{Duivenvoorden2016} as well as large scale environment such as proximity to a filament \citep{Malavasi2017,LaigleEtAl2017}. 

To help elucidate the mechanisms through which the environment affects galaxy and black hole evolution, we need surveys that reach high enough redshift to sample the epochs where the bulk of the stellar and black hole mass were assembled \citep[the bulk of stellar mass growth happened at $z\sim0.5-2$;][]{MadauDickinson2014}. We also need a large enough volume to sample a representative range of environments with good statistics. Spectroscopic surveys are ideal because of their ability to localize galaxies in 3D precisely; however, they do not reach to high enough redshift with high enough sampling rates \citep[e.g. see the VIPERS survey][for state of the art]{GuzzoEtAl2014} and tend to be biased against redder galaxies by selection. High quality photometric redshifts can work as shown for the COSMOS survey \citep[e.g][]{DarvishEtAl2015_filament,LaigleEtAl2016-cosmos2015}, but, with an area of only 2\,deg$^2$, this survey is not quite a representative cosmic volume and suffers from significant cosmic variance and poor statistics at the high-mass end \citep{Moster2011,DarvishEtAl2015,Yang2018}. 

The {\sl Spitzer} Extragalactic Representative Volume Survey \citep[SERVS; see][for survey definition and early results]{MauduitEtAl2012} was designed specifically to address these issues. With a total volume of $\approx 1$ Gpc$^3$ out to $z\sim3$, this survey reaches galaxies down to stellar masses of $M_*\sim10^{9.5}M_{\odot}$ at $z\sim 2$, corresponding to the epoch of ``cosmic noon'' and probes the full range of environments from voids to massive clusters.  
The survey centers on the {\sl Spitzer} IRAC data which samples the rest-frame near-IR out to cosmic noon and therefore allows for accurate stellar parameter estimation \citep{Muzzin2009}. The full multi-wavelength coverage, spanning from the X-rays to the radio, allows us to derive accurate photometric redshifts, star-formation rates and AGN presence and strength for our galaxies. In a series of papers, we use the SERVS and ancillary data to explore the role of environment in galaxy and black hole evolution. 

In this first paper of the series, we construct 2D density maps for the 4.8\,deg$^2$ XMM-LSS field where we have the most uniform and deep multiwavelength coverage in hand. The XMM-LSS field is already 2.5$\times$ the size of COSMOS with estimated $\approx$300 dark matter halos with log($M_{\rm halo}/M_{\odot})>$ 13.7.  The existing massive halo catalogs in this field are the X-ray cluster catalogs \citep{ClercEtAl2014,Adami2018}. Our density maps allow us to construct an independent and complementary catalog of the massive halos in this field. This is because of our catalog having a redshift-independent halo mass limit whereas X-ray cluster selection has a strong redshift dependence of its limiting mass. In addition, density-maps allow us to find overdensities that are not yet virialized, X-ray emitting halos. This paper is also a test case of what we can do with the quality of photometric data that are expected in the near future for a total of $\approx$15\,deg$^2$ spread across four fields with matching coverage from the $u$-band through the mid-IR. We demonstrate our ability to recover the highest density peaks and even pick up some large scale structure like filaments. We stress, however, that, being based on photometric redshifts, our overdensities are only candidates. They require spectroscopic confirmation. This should be available for many of these overdensities in the near future since this fields is also covered by the ongoing DEVILS and the upcoming  Prime Focus Spectrograph (PFS) spectroscopic surveys \citep{Davies2018,TanakaEtAl2017} which will significantly increase the spectroscopic coverage of the field out to cosmic noon.

This paper is organized as follows. In Section\,\ref{sec:data} we discuss the photometric and spectroscopic data in the XMM-LSS field. We also discuss the lightcone of simulated galaxies we use to help us interpret our observational results. In Section\,\ref{sec:analysis}, we present our analysis. This includes photometric redshift determination and uncertainty estimates, as well as density map generation in the simulated and observed dataset.  We present a list of 339 overdensity-selected cluster candidates and compare them with catalogs of spectroscopically-confirmed X-ray clusters in the field. In Section \ref{sec:conclusions}, we give our summary and conclusions. Throughout this paper, we adopt the Planck2015 cosmology \citep{Planck2015},  such that $\Omega_{m}$= 0.3075, $\Omega_{\Lambda}$= 0.691 and $H_{\rm 0} =  67.74$\,km\,s$^{-1}$Mpc$^{-1}$. All magnitudes are in the AB system. 

\section{Data \label{sec:data}}

\subsection{Photometry}

Our photometric data rely on the multiwavelength coverage of SERVS,  that provides {\sl Spitzer} IRAC 3.6 and 4.5\,$\mu$m data of a depth sufficient to reach below $M_*$ \citep[based on the compilation in][]{MadauDickinson2014} at cosmic noon across an area wide enough to cover a representative volume of the Universe. Specifically, it reaches a 5$\sigma$ point source depth of $\approx$2 $\mu$Jy (\emph{AB} = 23.1) and covers 18\,deg$^{2}$ spread across five fields to combat cosmic variance \citep[see][]{Moster2011}. Each field has an area of $\sim$2-5\,deg$^2$ which allows for large extended structures such as protoclusters and filamentary networks to be studied \citep{Yamada2012,Chiang2017}. 

\begin{figure*}
\centering
\includegraphics[width=0.45\textwidth]{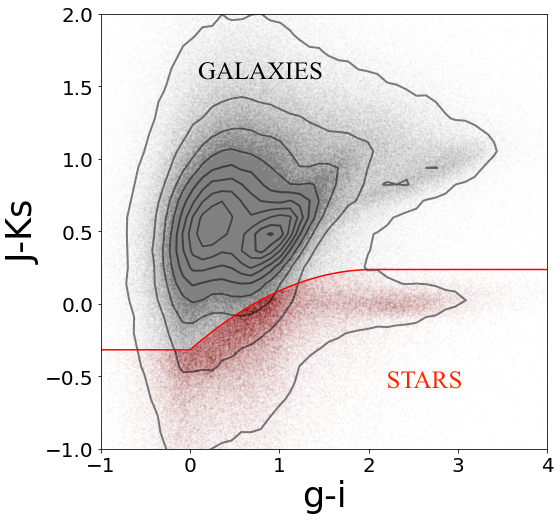}
\includegraphics[width=0.45\textwidth]{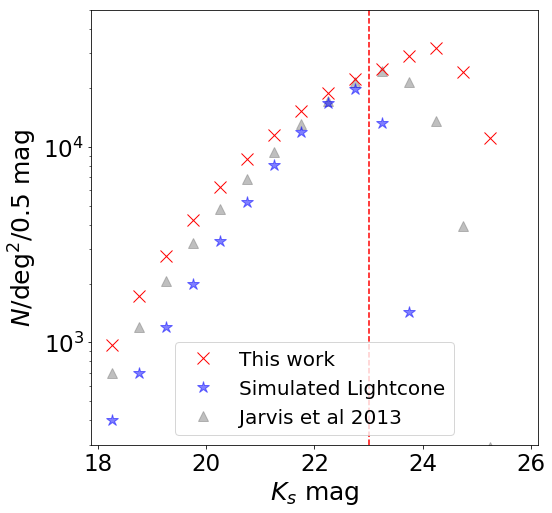}
\caption{{\it Left:} Separation of stars and galaxies using the \emph{J}-\emph{K$_{s}$} vs \emph{g}-\emph{i} color cut described in \citet{BaldryEtAl2010}. {\it Right:} The differential \emph{K$_{s}$} number counts after star removal. The counts from \citet{JarvisEtAl2013} are included for reference. The red dashed line indicates the \emph{K$_{s}$} magnitude limit of 23 which we adopt for this work for consistency with the simulated lightcone (note our lightcone is constructed based on a 3.6~$\mu$m-limited sample not a $Ks$-limited one).}
\label{fig:stars_counts}
\end{figure*}

In this paper we focus on the XMM-LSS field for which a new multi-band photometric catalog has been constructed using forced photometry \citep[][and Nyland et al. 2019 in prep]{NylandEtAl2017}.  This catalog was constructed using the Tractor code \citep{LangEtAlTractor2016} and uses one image as a reference for the source model which includes positions and surface brightness profiles of galaxies. The VIDEO $K_s$-band is the preferred reference image for the bulk of the galaxies, but other bands are used under certain circumstances such as gaps in coverage. These source models are applied across all other bands. This method is particularly crucial for deblending the IRAC 3.6 and 4.5\,$\mu$m photometry. We use this catalog for the $u$-through-4.5$\mu$m photometry. For the longer wavelength photometry including IRAC 5.8\,$\mu$m and 8.0\,$\mu$m and MIPS 24\,$\mu$m (from SWIRE; \citet{2003PASP..115..897L}) and {\sl Herschel} SPIRE 250, 350 and 500\,$\mu$m photometry we use the band-merged catalog published by the Herschel Extragalactic Legacy Program \citep[HELP;][]{Vaccari2016,Shirley2019}\footnote{http://hedam.lam.fr/HELP}. The HELP team also have published photometric redshifts in the field \citep{DuncanEtAl2018}, but this is based on their band-merged catalog (Shirley et al, in prep.) as opposed to the forced photometry catalog described above. We compare our photometric redshifts (derived below) with the HELP photometric redshifts. We find them to be consistent out to $z\sim1$; however, our IRAC-deblending (thanks to the forced photometry catalog) leads to more accurate redshifts at $z\sim1-2$. Therefore in this paper we use our own photometric redshift estimates. Since \citet{DuncanEtAl2018} use AGN templates in their photo-$z$ analysis, for the objects flagged as AGN (see below), we adopt the \citet{DuncanEtAl2018} photometric redshifts for AGN. We also use the spectroscopic redshifts and quality flags (see below) as compiled in the HELP catalog \footnote{See also: http://www.mattiavaccari.net/df/specz/}.

In the 4.8\,deg$^2$ XMM-LSS field, there are $\sim1.25$ million objects. We remove stars by using the \citet{BaldryEtAl2010} $J-K_{s}$ vs. $g-i$ color cut, leaving us with $\sim1.09$ million non-stellar objects (see Figure\,\ref{fig:stars_counts}). In the right-hand panel of Figure\,\ref{fig:stars_counts} we plot the differential \emph{K$_{s}$}-band number counts in the field using Tractor photometry and compare them with the counts using the VIDEO photometry from \cite{JarvisEtAl2013}. Tractor measures flux by fitting a surface brightness profile, whereas \citet{JarvisEtAl2013} use a fixed aperture. Tractor thus collects more flux from brighter, extended sources, accounting for the discrepancy between the two number counts at brighter magnitudes. We also overlay the counts from our simulated lightcone (see below for details). This lightcone assumes the original depth of SERVS in the IRAC bands \citep[see][]{MauduitEtAl2012} which leads to significant incompleteness above $K_s\sim23$. Our current Tractor photometry, which uses the $K_s$ image as a reference has allowed us to reach below the original single-band based photometry of SERVS. Since we rely on our simulated lightcone for analysis, though, we limit our sample to K$_{s}$ $<$ 23 to ensure a closer comparison. This yields 441,969 galaxies. We finally remove 16,211 additional objects classified as stars in SDSS (but missed in Figure\,\ref{fig:stars_counts}). This leads to a final sample of 425,758 galaxies.

In this sample we also flag any AGN, which we select based on: a) X-ray counterpart \citet[based on the][catalog]{ChenEtAl2018}, b) having AGN-like mid-IR colors, or c) being spectroscopically classified as AGN in SDSS, leading to a total of 2,113 potential AGN. AGN are not removed from the sample because we are interested in the effect of environment on the incidence, type, and strength of the AGN. However, given their typically significantly more uncertain redshifts we exclude them in the redshift quality assessment shown in Section\,\ref{sec:photoz} as well as the density map determinations in Section\,\ref{sec:analysis}.  Because the number of AGN is so small, this has negligible effect on our results.

\subsection{Spectroscopic data \label{sec:specz}}

The XMM-LSS field has significant spectroscopic redshift coverage. This is  primarily from the VIMOS VLT Deep Survey \citep[VVDS;][]{LefevreEtAl2013-vvds}
and the VIMOS Ultra-Deep Survey \citep[VUDS;][]{LefevreEtAl2014-vuds}. The VVDS and VUDS are \emph{i}-band magnitude selected surveys, going down to \emph{i}$_{AB}$ = 24 and \emph{i}$_{AB} \simeq 25$, respectively. These data are complemented by the VIMOS Public Extragalactic Redshift Survey (VIPERS) \citep{GuzzoEtAl2014} down to \emph{i}$_{AB}$ = 22.5, the PRIsm MUlti-object Survey (PRIMUS) \citep{CoilEtAl2011, CoolEtAl2013} down to \emph{i}$_{AB}$ = 23, the Sloan Digital Sky Survey (SDSS) \citep{AlamEtAl2015} down to \emph{i}$_{AB}$ = 21.3, the UKIDSS Ultra-Deep Survey (UDSz) \citep{BradshawEtAl2013,McLureEtAl2013} down to \emph{i}$_{AB}$ = 25 and the Australian Dark Energy Survey (OzDES) \citep{YuanEtAl2015} down to \emph{r}$_{AB}$ = 25. Note that these do not all cover the full area and targetted campaigns such as for cluster confirmation in the field \citep[see e.g.][]{Adami2018} are not included. These spectroscopic data are compiled as part of HELP and documented on their website\footnote{http://hedam.lam.fr/HELP}.
We also supplement with redshifts from the VANDELS survey \citep{McLure2018,Pantericci2018}, which uses the VIMOS spectrograph down to a limit of \emph{i}$_{AB}$=27.5. In total 76,016 sources in our sample have spectroscopic redshifts available in the XMM-LSS, 16,342 of which have been flagged as $>99\%$ reliable in the HELP catalog. This represents 3.7$\%$ of our sample.

\subsection{Simulated data}

To assess the reliability of environment measures in the presence of photometric redshift uncertainties, we employ a simulated catalog designed to cover an area and volume equivalent to the SERVS survey. Our simulated data are built from the Millenium N-body simulation \citep{Springel2005GADGET}. Model galaxies were constructed using the Lagos12 GALFORM semi-analytic model \citep{LagosEtAl2012,ColeEtAl2000-Galform} using the method described in \citet{MersonEtAl2013}. GALFORM models the main physical processes of galaxy formation and evolution, using the formation histories of dark matter haloes as a starting point. Our lightcone covers 18~deg$^2$ and spans the redshift range $0.0 < z < 6.0$, containing 1,518,854 galaxies\footnote{This lightcone is publicly avaialble at \url{astro.dur.ac.uk/~violeta/sw/servs.tar.gz}.}. This translates to $\approx$\,400,000 in an 4.8~deg$^2$. field consistent with our data (after the $Ks<23$ cut). Galaxy stellar mass and parent halo mass are both outputs of the lightcone. We also have simulated observations of each galaxy in the SDSS $z$-band, the DECam $Y$-band, and the UKIRT $J$-, $H$-, $K$-, and $Ks$-bands. The cosmology of the simulated lightcone is different than that which we assume for our observed sample, but we correct for this by multiplying masses by the value of $h$ appropriate to each sample.

\section{Analysis \label{sec:analysis}}
\subsection{Photometric redshifts \label{sec:photoz}}
Only $\sim$ 4\% of the sources in the field have spectroscopic redshifts (Section\,\ref{sec:specz}), so we need photometric redshifts in order to trace the galaxy density field. We determine photometric redshifts using the EAZY code \citep{Brammer-VD-Coppi-Eazy-2008}\footnote{There are many choices here. The fact that we see comparable $\sigma_{NMAD}$ and no significant systematic biases between our photometric redshifts and those derived for the same dataset but using different models by \citet{Pforr2019} suggest our results are likely robust against photometric redshift code systematics.}. EAZY compares input photometry to a linear combination of template spectral energy distributions (SEDs) realized across a range of redshifts, returning the redshift of the combination giving the smallest $\chi^{2}$ value and a redshift probability distribution function $p(z)$ for each galaxy. We use the EAZY template library including emission lines as well as a dusty star-forming template. We evaluate using the EAZY default redshift range from $0<z<8$ in steps of 0.01($1+z$). 

We employ an iterative zero-point correction algorithm \citep[see][]{BrodwinEtAl2006,IlbertEtAl2006} to correct for systematic magnitude deviations between the template set and our photometry. Here, we restrict ourselves to those galaxies with high quality spectroscopic redshifts for efficiency of iteration.\footnote{We tested using a random subsample of galaxies in our sample to avoid using the biased spectroscopic subsample. We found no significant differences in the resulting redshift distribution.} After determining photo-$z$'s for this subsample, we compute the median ratio between the best-fit template fluxes and the catalog fluxes in each photometric band. We then re-run EAZY, correcting the flux through each band by multiplying the flux by this ratio. We iterate this process until the flux ratios converge within 1\% of unity \citep[as in][]{IlbertEtAl2006}. The XMM-LSS is nearly uniformly covered by the SERVS survey (modulo a tiling pattern), but features three separate levels of depth in the optical from the HSC. 
Differences in coverage and depth affect the relative weight of given bands in the fit and therefore may affect the zero point corrections. We therefore perform this procedure separately for the HSC Wide, Deep, and UltraDeep patches. The zero-point magnitude offsets computed in this way are presented in Table\,\ref{tab:zpoffsets}. We note that these offsets are correcting for the the fact the particular template set used by EAZY may not be fully representative of the real galaxies spectra as well as for any systematic offsets in the photometry.

\begin{table}
\centering
\caption{Systematic offsets (mag$_{data}$ - mag$_{template}$) for each filter used to calculate photometric redshifts in the XMM-LSS catalog. }
\begin{tabular}{llll||llll}

\hline
\hline
Filter     & offset$_{Wide}$ & offset$_{Deep}$ & offset$_{UltraDeep}$   & Filter     & offset$_{Wide}$ & offset$_{Deep}$ & offset$_{UltraDeep} $ \\
\hline
\emph{u*} & 0.2748   &   0.2372  & 0.2472 & \emph{J} & 0.1235 & 0.0894  &0.1475  \\
\emph{g}   & -0.0031  &   -0.0848  &-0.0742  & \emph{H} & 0.0417 & 0.0024 &0.0370  \\
\emph{r}    & -0.0161     &  -0.2028  &-0.0610 & \emph{Ks} & -0.1102 & -0.1433  &-0.1131  \\
\emph{i}   & -0.0534  &    0.0440  & -0.0111  & [3.6] & 0.0207 & 0.0156  &0.0016 \\   
\emph{z} &  -0.0701    &   0.0662   & -0.0245 & [4.5] & -0.0029 & -0.0016  &-0.0031 \\
\emph{y}  &    0.0306   &     0.0337   &-0.0084 &      \\

\hline \hline
\end{tabular}
\label{tab:zpoffsets}

\end{table}

After the above procedure, the bulk of the sources ($>$96\%) show nominally excellent fits with reduced $\chi^2<3$. The $p(z)$'s are typically single-peaked, but show increasing incidences of multiple peaks with increasing $K_s$ magnitude. By adopting a $K_s<23$ cut as discussed above, we minimize the effect of multiple significant solutions. In addition, in the density map analysis presented in the next section, the full $p(z)$ profiles for each galaxy are taken into account and therefore any remaining sources with multiple peaks naturally have lower weight in the density map calculation.

\begin{figure*}
\centering
\includegraphics[width=0.42\textwidth]{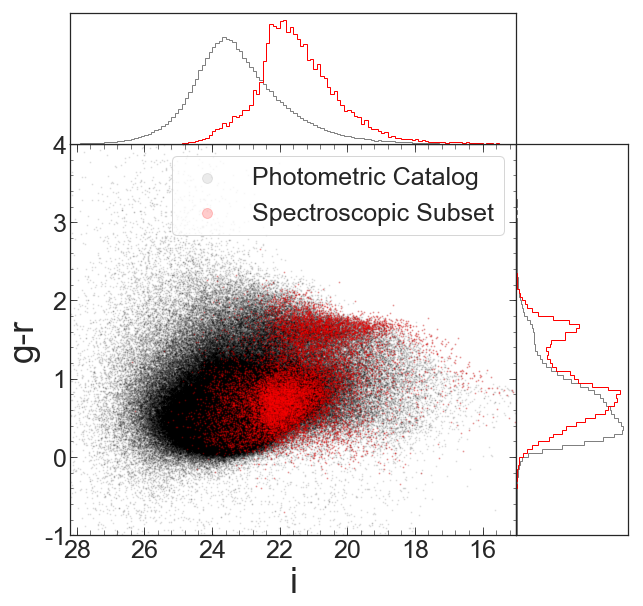}
\includegraphics[width=0.42\textwidth]{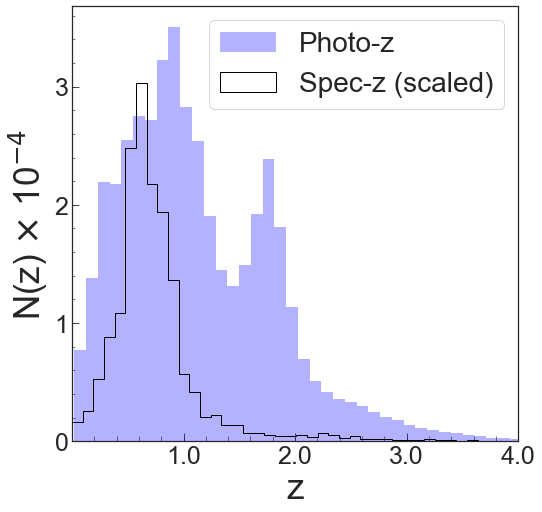}
\caption{{\it Left:} The \textit{g-r} color versus \textit{i}-band magnitude for our sample. The black points and histograms represent all galaxies in our sample, while the red points and histograms represent just those galaxies for which we have high quality spec-$z$'s as defined in the text. {\it Right:} The photo-$z$ distribution of our sample is shown (scaled as indicated) as the filled blue histogram. The spec-$z$ distribution (scaled for comparison) is plotted as the open histogram. 
\label{fig:photoz}}
\end{figure*}

Our choice of using a template-based method for deriving photometric redshifts is driven by the fact that non-template based methods rely more heavily on training on the spectroscopic sample. This requires a spectroscopic sample that is representative of the whole. The left-hand panel of Figure\,\ref{fig:photoz} shows the optical color-magnitude diagram comparing the full photometric catalog with the subset of galaxies with spectroscopic redshifts. It is clear that the spectroscopic subset is not representative of the whole. This is further highlighted in the right-hand panel of Figure\,\ref{fig:photoz}, where we show the photometric redshift distribution of all our sources compared with the spectroscopic redshift distribution. It is clear that at redshifts of $z \stackrel{>}{_{\sim}}1$, methods that heavily rely on spectroscopic redshift training will start to fail/be less reliable. We have not explored the validity of the observed peak in the redshift distribution at $z\sim1.9$, but our analysis does not extend that far therefore we ignore it at this point.  

\subsubsection{Photometric redshift accuracy}
\begin{figure*}
\centering
\includegraphics[width=0.33\textwidth]{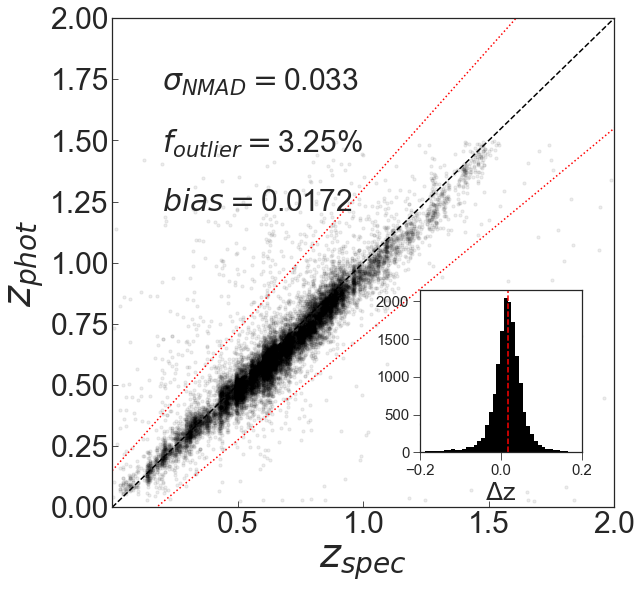}
\includegraphics[width=0.31\textwidth]{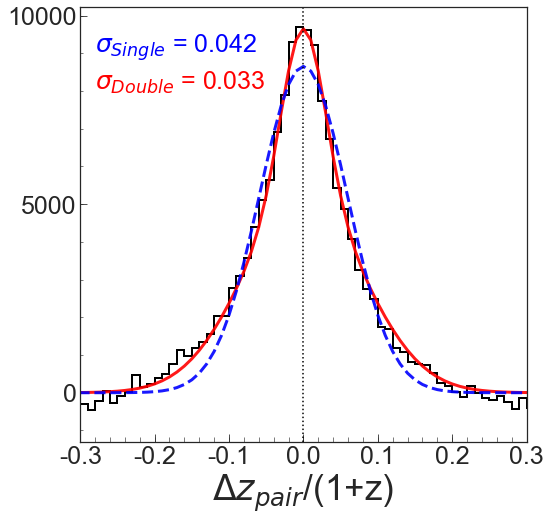}
\includegraphics[width=0.33\textwidth]{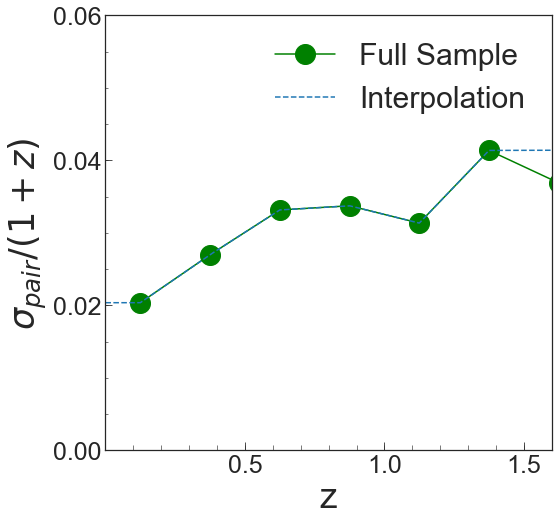}
\caption{{\it Left:} Comparison of photo-$z$ with quality spec-$z$ for galaxies with $z_{\rm phot}<1.5$. The inset shows the distribution of $\Delta z_{\rm pair}$ from which we derive $\sigma_{NMAD}$ after removing the outliers. Points that do not lie between the dotted red lines are considered outliers. {\it Middle:} Photo-$z$ uncertainty distribution for sources with $z_{\rm phot}<1.5$ based on the pair analysis. The single (blue) and double (red) Gaussian fits are overlaid. The double Gaussian is clearly a better fit to the distribution, and provides a weighted $\sigma_{\rm pair}/(1+z) = 0.033$. {\it Right:} Evolution of photo-$z$ uncertainty from $0.0<z<1.6$. The blue dashed line is an interpolation of the function as described in \ref{sec:slices}.}
\label{fig:zaccuracy}
\end{figure*}

We assess the photo-$z$ accuracy in two ways (Figure\,\ref{fig:zaccuracy}). We begin with the standard approach of comparing the photometric and spectroscopic redshifts.  We characterize the accuracy of the photo-$z$'s via: 
\begin{equation}
\sigma_{\rm NMAD} = 1.48 \times \rm{median} \frac{|z_{spec}-z_{phot}|}{1+z_{spec}},
\end{equation}
where $\sigma_{\rm NMAD}$ (the normalized median absolute deviation) is robust against catastrophic photo-$z$ failure. We identify outliers as objects with $|z_{\rm spec}-z_{\rm phot}|> 0.15$ \citep[as in][]{IlbertEtAl2013,DahlenEtAl2013}. Before computing $\sigma_{\rm NMAD}$, we remove outliers from the distribution. This helps characterize the distribution of only those sources which had relatively successful photo-$z$ fits. We calculate $\sigma_{\rm NMAD} = 0.033$ for our field, with outlier fraction $f_{\rm outlier} = 3.25\%$ and a bias of 0.0172. We correct for this bias when we compute our density maps. 

However, this standard approach has the drawback that the sub-sample with spectroscopic redshifts is not representative of the sample as a whole in a color-magnitude diagram (Figure\,\ref{fig:photoz}). To overcome this, we also characterize our uncertainty using the pair method of \citet{Quadri2010}. This method does not give reliable outlier fractions or bias estimates, but uses the full photometric sample to compute $\sigma_z/(1+z)$. This gives us sufficiently large numbers, especially above $z\sim1$, to allow us to assess the uncertainties as a function of redshift. 

The pair method exploits the observation that galaxies with close angular separation have a significant probability of being physically associated, while the role of line-of-sight projections can be subtracted in a statistical sense. It works as follows. For each galaxy in our sample, we first search for pairs separated by 2.5-15\arcsec. Galaxies often contain multiple pairs in this annulus, so the following procedure is done for all such pairs. For each pair we compute:
\begin{equation}
    \Delta z_{\rm pair} = (z_{\rm phot,1} - z_{\rm phot,2})/(1+ z_{\rm mean}) ,
    \label{eqn:delta_zpair}
\end{equation}
where $z_{\rm mean}$ represents the mean redshift of the pair. We then randomly assign each object new coordinates in the field and perform the same procedure to obtain the distribution of $\Delta z_{\rm pair}$'s for random line-of-sight projections. Subtracting the random distribution of $\Delta z_{\rm pair}$'s from the observed distribution of $\Delta z_{\rm pair}$'s gives a the distribution for physically associated galaxy pairs only\footnote{The characteristic velocity differences between such physical pairs are much smaller than the photometric redshift uncertainties. Therefore, we ignore this effect.}. This distribution for our full sample of $z<1.5$ galaxies is shown in the middle panel of Figure\,\ref{fig:zaccuracy}. The width of this distribution is a factor of $\sqrt{2}$ larger than the photo-$z$ uncertainty per galaxy (because we are dealing with pairs of galaxies).

In the middle panel of Figure\,\ref{fig:zaccuracy}, we show that a single Gaussian can be a poor fit due to the presence of extended wings on either side of the peak. As discussed in \citet{Quadri2010}, these wings likely result from the non-flat redshift distribution of our field. We follow the recommendation of \citet{Quadri2010} and fit the distribution by a convolution of two Gaussians \citep[see][for full details]{Quadri2010} centered about zero. The weighted sum of the standard deviations of the two Gaussians gives $\sigma_{\rm pair}/(1+z)$ using this method. For our sample, this double Gaussian is clearly the better fit, and gives an uncertainty $\sigma_{\rm pair}/(1+z) = 0.033$, consistent with $\sigma_{\rm NMAD}$ derived above. 

The right-hand panel of Figure\,\ref{fig:zaccuracy} shows the thus derived uncertainty as a function of redshift, in bins of $\Delta z = 0.25$. We find the uncertainties $\sigma_{\rm pair} / (1+z)$ get significantly worse at $z\gtrsim1.5$; therefore, for the density map analysis in the following sections, we will restrict ourselves to $0.1 < z<1.5$.
%\footnote{}

\begin{table}
\centering
\caption{Redshift slices and comoving depths }
\begin{tabular}{ccc||ccc}

\hline
\hline
bin & $z$-range     & depth (Mpc) & bin  & $z$-range   & depth (Mpc) \\
\hline
1 &  0.1 $<$ z $<$ 0.146 & 192   &   15 & 0.549 $<$ z $<$ 0.649 & 317 \\
2 & 0.123 $<$ z $<$ 0.171 & 197   &   16 & 0.599 $<$ z $<$ 0.705 & 324 \\
3 & 0.146 $<$ z $<$ 0.196 & 203   &   17 & 0.649 $<$ z $<$ 0.760 & 330  \\
4 & 0.171 $<$ z $<$ 0.224 & 210  &   18 & 0.705 $<$ z $<$ 0.820 & 330   \\
5 & 0.196 $<$ z $<$ 0.251 & 217  &   19 & 0.760 $<$ z $<$ 0.879 & 330  \\
6 & 0.224 $<$ z $<$ 0.282 & 227   &   20 & 0.820 $<$ z $<$ 0.942 & 325   \\
7 & 0.251 $<$ z $<$ 0.313 & 237  &   21 & 0.879 $<$ z $<$ 1.004 & 322  \\
8 & 0.282 $<$ z $<$ 0.348 & 246   &   22 & 0.942 $<$ z $<$ 1.069 & 315   \\
9 & 0.313 $<$ z $<$ 0.382 & 254  &   23 & 1.004 $<$ z $<$ 1.133 & 309  \\
10 & 0.348 $<$ z $<$ 0.421 & 265 &   24 &  1.069 $<$ z $<$ 1.207 & 318   \\
11 & 0.382 $<$ z $<$ 0.460 & 275  &   25 & 1.133 $<$ z $<$ 1.281 & 327  \\
12 & 0.421 $<$ z $<$ 0.504 & 287   &   26 & 1.207 $<$ z $<$ 1.372 & 347   \\
13 & 0.460 $<$ z $<$ 0.549 & 298  &   27 & 1.281 $<$ z $<$ 1.462 & 365  \\
14 & 0.504 $<$ z $<$ 0.599 & 308   &   28 & 1.372 $<$ z $<$ 1.564 & 368   \\

\hline \hline
\end{tabular}
\label{tab:zbins}

\end{table}

\subsection{Surface density maps: method and validation}
We follow the method of \citet{DarvishEtAl2015}, who construct 2D density maps based on photometric (and, where available, spectroscopic) redshifts in the COSMOS field. In this method the surface density is estimated in redshift slices via weighted adaptive kernel smoothing. This means that the kernel width adapts to the local density as described below. 
This method allows us to probe adaptively smaller volumes than the more commonly-used fixed aperture `cylinders'. The weights account for photometric redshift uncertainties which helps combat the significant smearing of the signal along the line of sight 
\citep[see][for a comparison between different density estimation methods]{MuldrewEtAl2012}. 

\subsubsection{Redshift slices}
\label{sec:slices}
To obtain our redshift slices, we first linearly interpolate the redshift evolution of $\sigma_{\rm pair}/(1+z)$ onto a finer grid. Starting with $z=0.1$ we set the start and end points for our bins such that each slice width is $2\times\sigma_{\rm pair}$ evaluated using the interpolation shown in the right-hand panel of Figure\,\ref{fig:zaccuracy}. We use overlapping redshift slices so that if a given overdensity falls on the edge in one slice and therefore is significantly scattered outside of the slice it will, by design, end up in the middle of the neighboring overlapping slice. Our redshift slices and their co-moving depths are given in Table \ref{tab:zbins}.

\subsubsection{Weighted adaptive kernel smoothing}
The weighted adaptive kernel smoothing works as follows. For each redshift slice, we identify the objects whose median photo-$z$ falls within that slice. We then weight each of the galaxies within our redshift slice by the fraction of that object's total $p(z)$ that lies within the slice of interest. We proceed to estimate the surface density $\hat{\Sigma}$(\textbf{r}$_{i}$) at each object's location by summing over a kernel \textit{K}. In \citet{DarvishEtAl2015} this kernel is a weighted 2D Gaussian whose width starts at $h=0.5\,$Mpc, but is adaptive, i.e. scaled by the local density in a manner analogous to that described below. We test this algorithm first by taking the public COSMOS data and reproducing the density map around a filament at $z\sim0.5$ that was published in \citet{DarvishEtAl2015_filament}. While we successfully reproduce the results of that study, we find the code to be slow\footnote{We ran timing tests and found the full Gaussian kernel to be 25\,$\times$ slower for a test run of $\approx10^3$ objects compared to the alternatives listed, which all took comparable time.}. We test three faster alternatives: a truncated Gaussian kernel $K_{G}(h)$, an Epanechnikov (parabolic) kernel $K_{E}(h)$, and a top-hat kernel $K_{T}(h)$. The Epanechnikov kernel is defined as:
\begin{equation}
K_{E}(\textbf{r}_{i},\textbf{r}_{j},h) = \left \{     \begin{array}{ll}
    \frac{3}{4h^{2}}(1-(\frac{r_{i} - r_{j}}{h})^{2}) & \textup{where } |r_{i} - r_{j}| < h \\ 
    0 & \textup{else,}
\end{array}\right.\\
\label{eqn:epanechnikov-kernel}
\end{equation}
where \textbf{r}$_{i}$ is the position of the object and \textbf{r}$_{j}$ is the position of each other object.

Following \citet{DarvishEtAl2015}, we choose an initial fixed kernel width \textit{h} for the Gaussian kernel of 0.5\,Mpc, which corresponds roughly to $R_{200}$ for a halo of 10$^{13}$ M$_{\odot}$. For the Epanechnikov kernel we get equivalent results using an initial value of $h=1.0$ Mpc. Using a fixed width would underestimate the density in overdense regions and overestimate the density in underdense regions, so we calculate an adaptive smoothing width $h_i = h \times \lambda_{i}$ for each object. Here $\lambda_{i}= \sqrt{G/\hat{\Sigma}(\textbf{r}_{i})}$, where \textit{G} is the geometric mean of all $\hat{\Sigma}$(\textbf{r}$_{i}$) values. To compute the surface density in our redshift bins, we set up a regular grid in steps of 50\,kpc. Thus, 50\,kpc sets the minimum scale probed by our maps in the plane of the sky. On top of this grid, the density maps are computed using the same kernel, but now using the adaptive width $h_{i}$. This surface density is converted to an overdensity with respect to the median surface density $\Sigma_{\rm m}$ in the slice as follows:
\begin{equation}
\delta = \frac{\Sigma - \Sigma_{\rm m}}{\Sigma_{\rm m}}.
\end{equation}

\begin{figure*}
\centering
\includegraphics[width=0.9\textwidth]{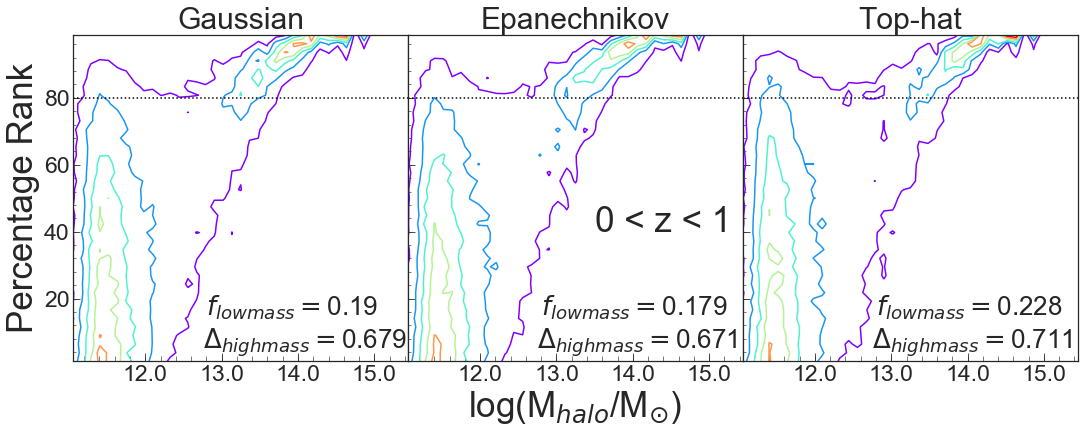}
\includegraphics[width=0.9\textwidth]{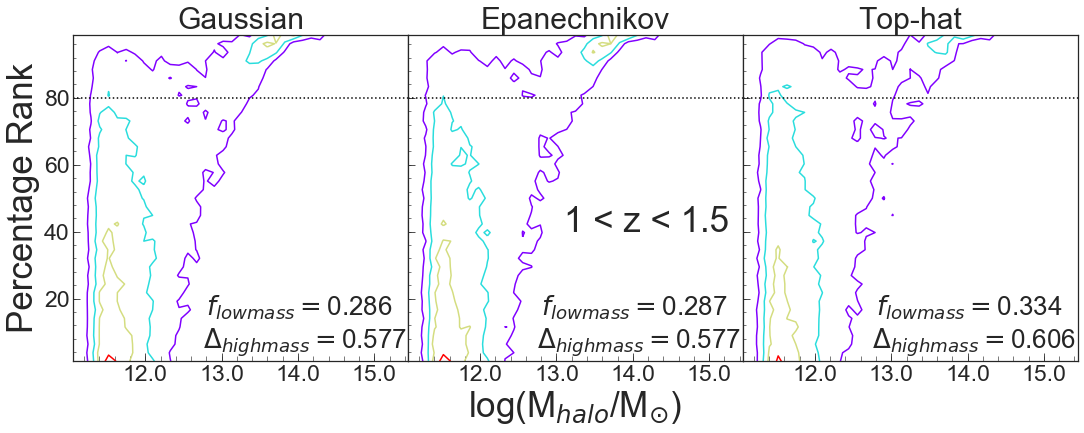}
\caption{Here we use the 2D density maps calculated using the exact redshifts of the simulated galaxies to show how the overdensity percentile compares with the host halo mass associated with each simulated galaxy. The top panel uses all galaxies in the range 0$<$\,$z$\,$<$1 and the bottom ones uses all galaxies in the range 1$<$\,$z$\,$<$1.5. Contours are linearly spaced and indicate constant galaxy number. These show that the bulk of galaxies that reside in high ($>80^{\rm th}$) percentile overdensities live in high mass halos (log($M_{\rm halo}/M_{\odot})>$13 which is roughly group-scale). Therefore finding high percentile overdensities helps us find high mass halos.   We compare this behavior using a truncated Gaussian filter (\emph{left}); an Epanechnikov kernel (\emph{middle}); and a top-hat kernel (\emph{right}). Here $f_{\rm lowmass}$ is the fraction of galaxies above the $80^{\rm th}$ percentile within log$(M_{\rm halo}/M_{\odot})<12$, while $\Delta_{\rm highmass}$ is the width of the distribution of galaxies above the $80^{\rm th}$ percentile. On both counts, the Epanechnikov filter performs marginally better than the truncated Gaussian; while the top-hat filter is the worst-performing.
\label{fig:perc_rank_comparison}}
\end{figure*}

To help us decide between the different kernel options, we compute density maps using exact redshifts from the simulated lightcone and the bins defined in Table \ref{tab:zbins}. In Figure\,\ref{fig:perc_rank_comparison}\,(top panels) we show the relation between the percentage rank of measured 2D overdensities and dark matter halo mass (using the main host halo mass for each galaxy in our simulated lightcone, not sub-halos) for all galaxies in the range $0<z<1$. For all kernels, we find that the most overdense regions tend to correlate with high halo masses, suggesting we can indeed recover the peaks of the density field with this method. To further help us differentiate between them, we consider two measures. The first tells us how often a significant overdensity is observed when there is no corresponding high mass halo (i.e.\ a false positive measure). We compute this as the fraction of galaxies found in $>$80$^{\rm th}$ percentile overdensities that are located in halos with log($M_{\rm halo}/M_{\odot})<$12 ($f_{\rm lowmass}$).  The other important measure is how closely halo mass maps onto overdensity percentile (i.e. an accuracy measure). We compute this as the width of the halo mass distribution for all galaxies found in 80th percentile overdensities after removing the galaxies in halos with log($M_{\rm halo}/M_{\odot})<$12 ($\Delta_{\rm highmass}$). The Epanechnikov filter performs the best in both measures, although the differences between the three filters are fairly minor. In the bottom panels of Figure\,\ref{fig:perc_rank_comparison}, we do the same analysis for galaxies in the range $1<z<1.5$. We find that the basic trends hold although, as expected, the false positive rate is now higher. This however is the result of using a fixed percentile (here 80\%) in selecting significant overdensities. As discussed below, we find it is more appropriate to use an evolving percentile threshold.   

\begin{figure*}
\centering
\includegraphics[width=0.9\textwidth]{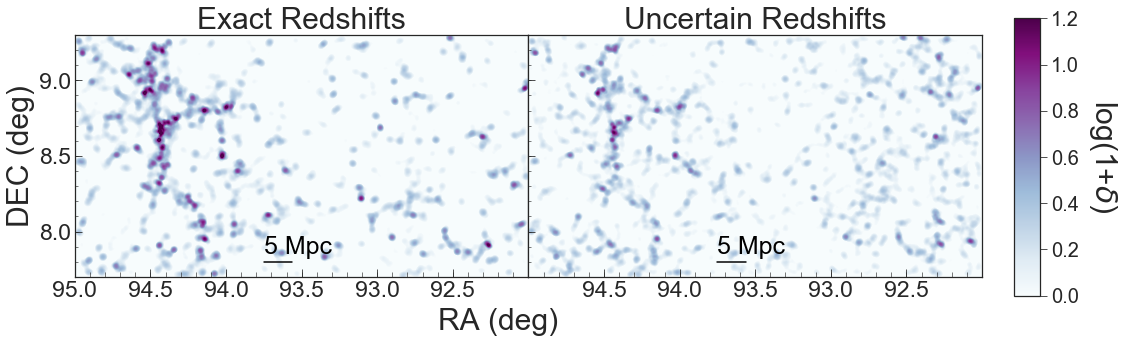}
\caption{Smoothed density maps of our simulation over the range $0.649<z<0.760$ using the Epanechnikov kernel. \emph{Left:} Density map using exact redshifts. \emph{Right:} Density map using uncertain redshifts mimicking our photometric redshift errors as described in text. These show that while the high density peaks are weaker in the uncertain redshifts map, overall both the peaks and the large scale structure are preserved.}
\label{fig:simulated_maps}
\end{figure*}

\subsubsection{Probing the density field with photometric redshifts \label{sec:recoverability}}

Here we investigate the recoverability of true overdensities given the uncertainties in our photometric redshifts. We choose an arbitrary 4.8~deg$^2$ subsection of the simulation to match the area of our XMM-LSS field and perform the density map analysis using both the exact redshifts and redshifts minimicking the photometric redshifts (see Figure\,\ref{fig:simulated_maps}). To mimic what we do with the observed data, where we use the $p(z)$ output from EAZY for each galaxy, we assign an uncertain ``photometric" redshift to each galaxy by drawing from a Gaussian distribution centered on the galaxy's exact simulated redshift $z_{\rm exact}$ and a width given by the typical width of a galaxy's $p(z)$\footnote{We do not specifically simulate double peaked distributions as these form a negligible fraction of the whole for our $K<23$ sample.}. We note that this is wider than $\sigma_{\rm pair}(z)$ since this incorporates outliers, multiple-$z$ solutions etc. We use this wider width as it approximates better what we can do with the data itself. We know that this works since we get reasonably close distributions of ${\rm log}(1+\delta)$ between simulations and data in this way.  For a random subset of 3.25$\%$ of the objects, we assign a redshift drawn from a uniform distribution in the range $0<z<3$ to mimic the catastrophic outlier fraction. Lastly, 3.7$\%$ of the objects are given redshifts equal to their exact simulated redshift to mimic our spec-$z$ fraction. The latter fraction of course is a function of redshift. However, it is small enough that its role is negligible here. 

Figure\,\ref{fig:simulated_maps} shows an arbitrarily chosen redshift slice at 0.649 $< z < $ 0.760, where we compare the surface density computed using the exact redshifts vs. uncertain redshifts as determined above. This figure highlights that the main large-scale structures and strongest overdensities are largely preserved. This means that we are likely to recover features, real or projected, that would appear if we had exact redshifts given the slice width we are using. There are, however, significant differences as well. The amplitudes of the overdensities are generally lower in the uncertain redshifts map (as expected). This is because the uncertainty in the redshifts tends to `smear' the signal and decrease the magnitude of over- and underdensities. Also apparent are some weaker ``false" features in the uncertain redshifts map not present in the exact map (see e.g. upper-right end of the maps). These arise due to real overdensities that belong to neighbouring redshift slices being scattered into this slice by redshift uncertainties. 

\begin{figure*}
\centering
\includegraphics[width=0.3\textwidth]{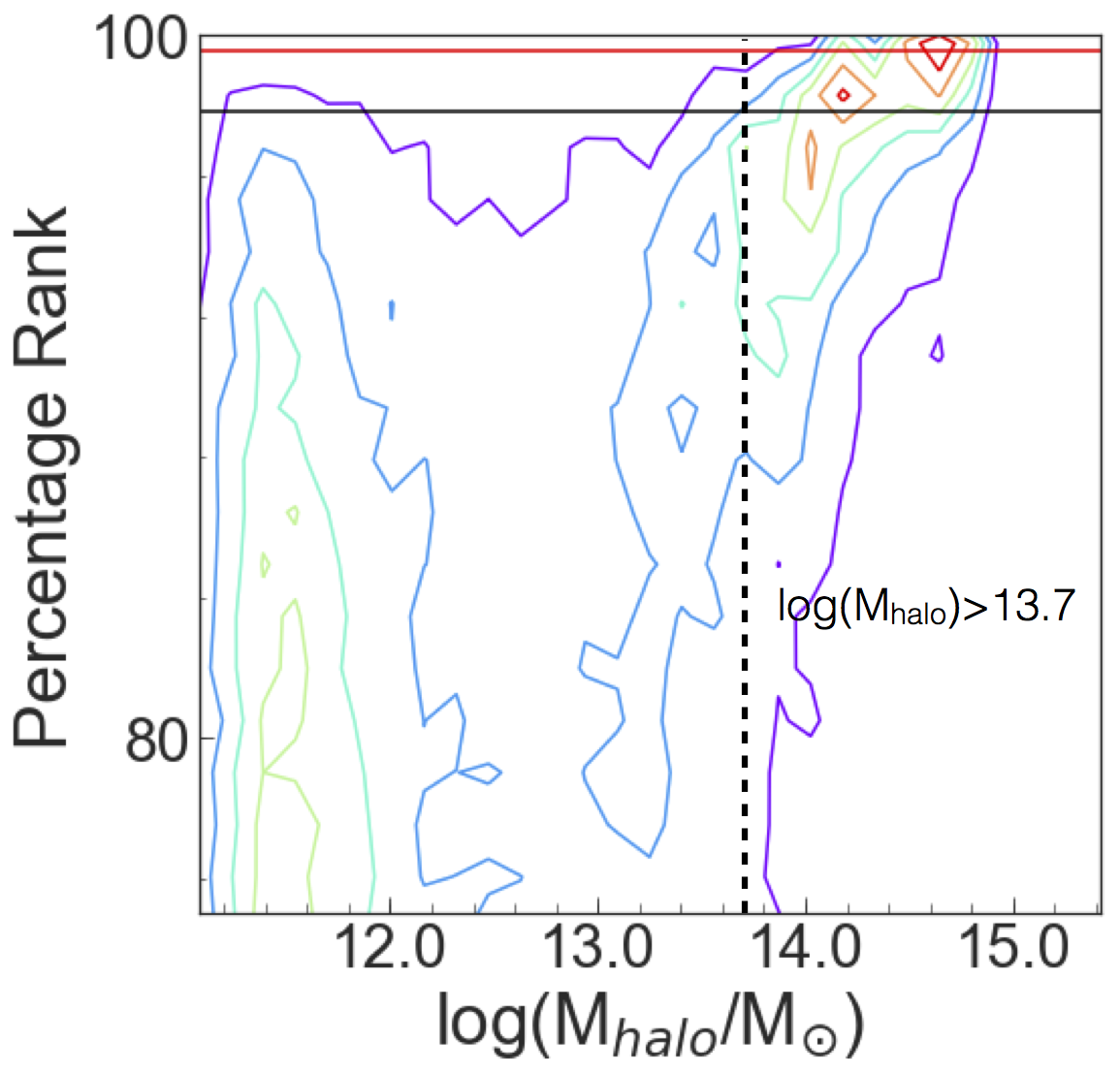}
\includegraphics[width=0.3\textwidth]{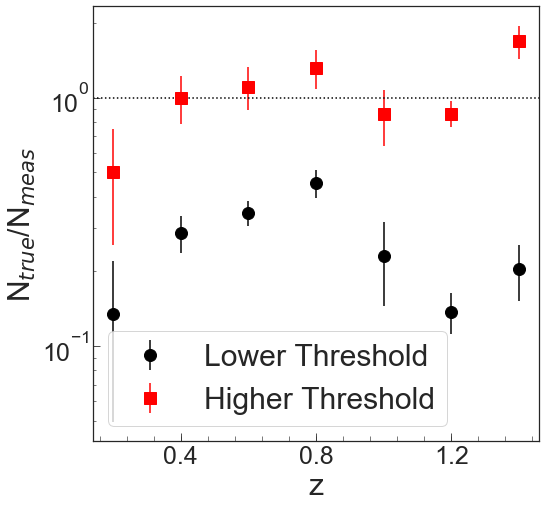}
\includegraphics[width=0.32\textwidth]{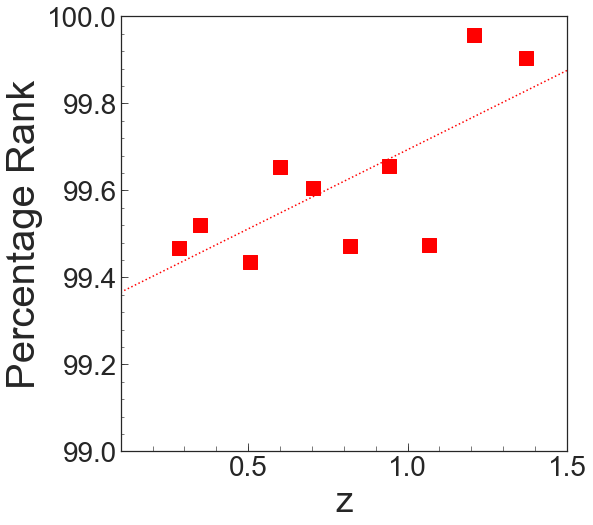}
\caption{ {\it Left:} Similar to Figure\,\ref{fig:perc_rank_comparison}, but now including the uncertainty in the photometric redshifts. This has the effect of scattering galaxies from low density environments to high density ones and vice versa. The group scale is no longer reliably recovered as many galaxies in smaller mass halos are found in $>80\%$ overdensities. If we consider the recoverability of cluster scales (here defined $M_{\rm halo}>10^{13.7}$M$_{\odot}$) however, they can still be recovered at higher percentile levels. The red and black horizontal lines are two test recoverability percentiles. {\it Middle:} The recoverability of log($M_{\rm halo}/M_{\odot})>$ 13.7 using these two test thresholds as evaluated by $N_{\rm true}/N_{\rm measured}$ (see text for details). The dotted black line corresponds to $N_{\rm true}/N_{\rm measured}= 1$ which corresponds to a minimal false positive rate -- the higher red threshold is therefore preferable by this measure. {\it Right:} Median percentage ranks to recover log(M$_{\rm halo}$/M$_{\odot}$)$>$ 13.7 as a function of redshift. Note that this threshold increases with redshift as expected from hierarchical halo growth.
\label{fig:overdensity_recoverability}}
\end{figure*}

Next we investigate how reliably 2D density maps calculated including redshift uncertainties track the mass of the host halo of the galaxies in the simulated lightcone. The left-hand panel of Figure\,\ref{fig:overdensity_recoverability} is similar to Figure\,\ref{fig:perc_rank_comparison} in that it plots the percentile of the overdensity in which a galaxy is found against its host halo mass. However, here the density maps are computed using uncertain redshifts (mimicking our photometric redshift errors). We again see a bimodality in this distribution about a group scale (log($M_{\rm halo}/M_{\odot})\sim$13), above which halo mass correlates with observed 2D overdensity. However there are two marked differences: 1) there is significant presence of lower mass halos in high density parts of the maps; and 2) there is significant range in observed percentile per halo mass. Given the first point, we can only reliably recover halos with log($M_{\rm halo}/M_{\odot})\gtrsim$ 13.7) going to higher percentiles in the density maps. 

How reliably can we recover clusters (here defined as log($M_{\rm halo}/M_{\odot})>$ 13.7)? Given the second point above, we need to find what percentile threshold  mass for finding halos above said scale, minimizes false positives and false negatives. The middle panel of Figure\,\ref{fig:overdensity_recoverability} illustrates this further by considering different two test thresholds aimed at finding halos with log($M_{\rm halo}/M_{\odot})>$13.7. Here we measure
$N_{\rm true}/N_{\rm measured}$, the ratio of all halos in the simulation that meet the mass criteria, to all such halos that are recovered using the particular percentile threshold. This ratio ideally should be $\approx$1 with a value $<<1$ showing significant number of false positives (suggesting the percentile threshold is too low) while a value $>>1$ shows significant number of false negatives suggesting the percentile threshold is too high). 

In the right-hand panel of Figure\,\ref{fig:overdensity_recoverability} we show the percentile threshold that recovers log($M_{\rm halo}/M_{\odot})>$ 13.7 with $N_{\rm true}/N_{\rm measured}\approx 1$ as a function of redshift. This threshold increases with redshift as expected from hierarchical halo mass growth -- i.e. a high mass halo is more of an extreme in the density field at higher redshift than at lower redshift. This evolving threshold is given by:

\begin{equation}
\centering
    Percentile_{13.7} = 0.355z + 99.4 .
\label{eqn:thresholds_halomass}
\end{equation}

\begin{figure*}
\centering
\includegraphics[width=0.7\textwidth]{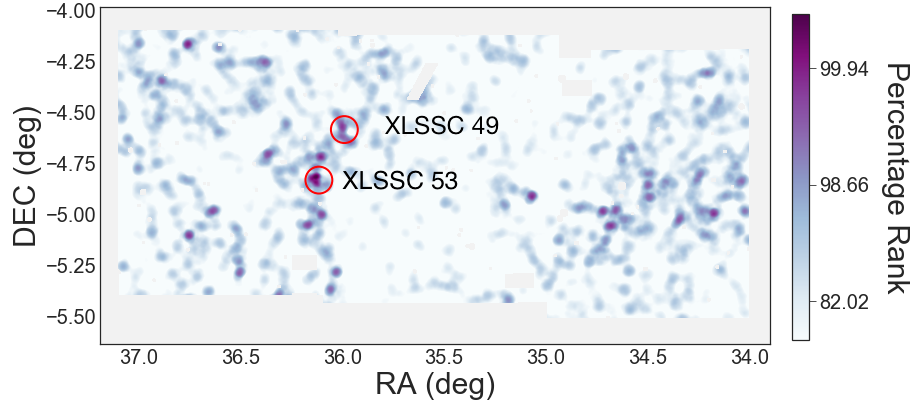}
\caption{A sample XMM-LSS density map at 0.421$<z<$0.504. The red circles mark two known spectroscopically-confirmed X-ray clusters \citep{ClercEtAl2014}, XLSSC 49 and 53, both of which are located at z=0.50. }
\label{fig:density-maps-xmm}
\end{figure*}

\subsection{Observed SERVS XMM-LSS density maps}
  
To compute the surface density maps for our observed data in the XMM-LSS field, we first correct for the measured bias in the photometric redshifts by multiplying the redshifts by 1.0172. We also generate a mask where bright stars (based on the bright star mask of \citet{MauduitEtAl2012}), the field border, and image artifacts are all masked. This mask brings the usable area of the field down to 4.0~deg$^2$ For consistency, we mask a 0.8~deg$^2$ area out of the border of the simulated field. This matches the area though obviously not the exact geometry of the observed masked field -- this is not relevant to the present paper though needs to be considered for any large scale structure studies.

In addition, there is some non-uniformity of coverage across the field, in particular with about 1/3 being covered by the deeper optical HSC-UltraDeep patch, and the rest being covered by HSC-Deep. The difference in depth in the $grizy$ bands leads to lower uncertainties in the photometric redshifts for the HSC-Ultradeep patch than in the HSC-Deep patch. This means they suffer less smearing and we end up seeing higher overdensities of sources in this patch especially at $z\gtrsim1$. We found that this non-uniformity is minimized by using percentiles rather than the measured overdensities in determining thresholds for finding significant overdensities (which is why we used percentiles in finding the threshold for selecting ${\rm log}(M_{\rm halo}/M_{\odot})>13.7$ halos in Section\,\ref{sec:recoverability}).

Figure\,\ref{fig:density-maps-xmm} gives an example of a density map for the XMM-LSS field. It shows qualitatively similar behavior to that of the simulated maps shown in Figure\,\ref{fig:simulated_maps}. In particular, we see very similar large scale filamentary structures. We defer our analysis of such structures to a subsequent paper in this series. We also circle two previously known spectroscopically confirmed X-ray clusters, both recovered by our density maps (Overdensities 69 and 92 in Table\,\ref{tab:cluster_candidates}, see Section\,\ref{sec:newClusters} for details).

\begin{figure*}[htb!]
  \centering
  \includegraphics[width=.25\textwidth]{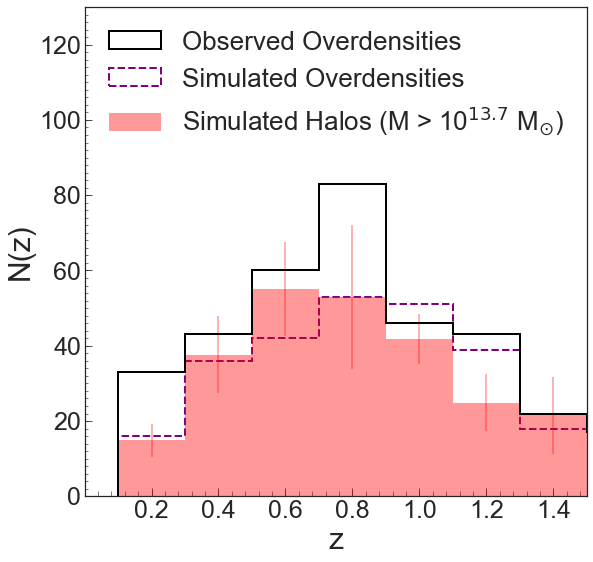}
   \includegraphics[width=.36\textwidth]{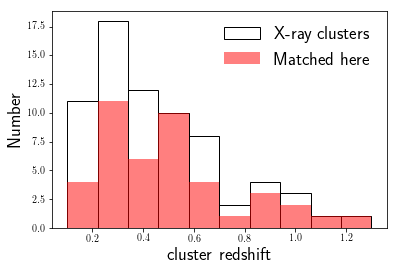}
   \includegraphics[width=.37\textwidth]{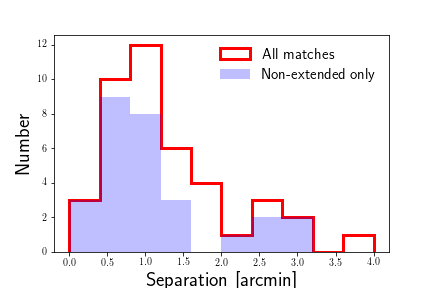}
  \caption{
 {\it Left:} Histogram of overdensities corresponding to M$_{\rm halo} \gtrsim$ 10$^{13.7}$ within an area of 4.0~deg$^2$. The solid black line are the distinct overdense regions above the threshold in our sample. In purple are the distinct overdense regions above the threshold in the simulated lightcone with degraded redshift accuracy. For reference, in red is the distribution of halos with M$_{\rm halo} \gtrsim$ 10$^{13.7}$ in the simulation. in our sample as a function of redshift. The height of the red histogram is found by dividing the 18~deg$^2$ of the lightcone into four quadrants and averaging the count of halos in each quadrant (scaled by 4.0/4.5). The error bars represent the maximum and minimum count among the four quadrants to give some sense of the impact of cosmic variance.
  {\it Middle:} The open histogram represents the spectroscopically-confirmed X-ray clusters in the field based on \citet{Adami2018}, where the minimum cluster mass is ${\rm log}(M_{500}/M_{\odot})=13.3$ which is below our target halo mass. The X-ray selection leads to increasing mass limit with increasing redshift
  The filled histogram represents the known X-ray clusters we recover from our density maps. The relative dearth of recovered clusters at lower redshifts is the result of the lower cluster masses thereof. {\it Right:} The separations between the positions of our overdensities and their closest matching X-ray clusters. Note that at $z\sim0.2-1$ 1\,arcminute corresponds to $\approx$200-500\,kpc. The shaded histogram shows this distribution for the non-extended overdensities only. It is clear that these separations are largely driven by the uncertainty in the center positions of the overdensities.
  \label{fig:cluster_recoverability}}
\end{figure*}

\subsection{Potential new clusters in XMM-LSS \label{sec:newClusters}}
Using our density maps for XMM-LSS and the percentile threshold for recovering halos of ${\rm log}(M_{\rm halo}/M_{\odot})>13.7$ given in Equation\,\ref{eqn:thresholds_halomass}, we find 330 potential halos with ${\rm log}(M/M_{\odot})>13.7$ between $0.1 < z < 1.5$ (339 if we allow for a few more clusters that may be in this range or in the slice just above).  Due to the use of overlapping redshift slices, overdensities found in neighbouring slices in overlapping RA-Dec locations are counted only once. They are assigned to the slice in which they are strongest.  Our overdensities are listed along with their basic characteristics in Table\,\ref{tab:cluster_candidates}. Note that for overdensities selected in more than one slice the redshift ranges given in the table conservatively span both slices.

The redshift distribution of our overdensities is given in the left-hand panel of  Figure\,\ref{fig:cluster_recoverability}, where overdensities spanning more than one redshift slice (see above) are assigned to the one where they have the highest percentile. We overlay on that the redshift distribution of all halos with log($M_{\rm halo}/M_{\odot})>$13.7 expected in our field based on our simulated lightcone. We expect 279$\pm$35 such halos in the XMM-LSS. This errorbar represents the spread in this number when drawing random 4\,deg$^2$ patches from our 18\,deg$^2$ simulated lightcone and therefore approximates cosmic variance\footnote{This is only a rough approximation since in our lightcone we have $<$5 independent realizations.}. We note that, at nearly 13\%, cosmic variance is significant at this mass scale even for our relatively large field. On the same figure we overlay the histogram derived by applying photometric redshift uncertainties to our lightcone, generating 2D density maps and extracting likely clusters based the exact same procedure as applied to the real data. The fact that this distribution recovers quite closely the simulated halos histogram, suggests our procedure is robust in recovering such structures\footnote{Since we ultimately derive our percentile threshold on the same simulations, this is not surprising but a good sanity check. Of course, we also assume that the simulated lightcone itself is a good representation of reality which previous studies suggest is the case \citep[e.g.][]{LagosEtAl2012}.}. The recovered such structures from the observed XMM-LSS field is also quite close -- indeed it is only $\approx18\%$ above the expected number from the simulated lightcone (only a few percent above if we take the upper limit from our cosmic variance test). This difference is small enough that it can still be attributed to cosmic variance (given our rough estimate thereof). However, it can also be due to a non-negligible false positive rate. This could result from slightly overestimating our redshift uncertainties -- which means we have less redshift smearing and somewhat higher densities than in the simulations.  

In Table\,\ref{tab:cluster_candidates} we also include a morphology column which flags overdensities that have significant spatial extent. This is computed by making 80$\times$80 pixel cutouts around each overdensity and computing the fraction of that patch that is above our density threshold. By visual inspection we chose $>10\%$ as signifying extended overdensities. About 1/4 of our overdensities fall in this category. These are likely to be dominated by projection effects given the wide redshift slices as well as the effects of large scale structure such as clusters apparently embedded in filaments as shown in Figure\,\ref{fig:density-maps-xmm}. These can also include cases of cluster mergers as discussed in Section\,\ref{sec:case_studies}. 

\subsubsection{Confirming previously known X-ray clusters}
In Table\,\ref{tab:cluster_candidates} we also note which of our overdensities are previously known in the literature. For this comparison we looked at the XLSSC catalog of \citet{Adami2018}. Within our field and above $z=0.1$, there are 70 X-ray clusters that are spectroscopically confirmed from this catalog. We matched these against our overdensities using a 0.07~deg separation (corresponding to $\approx$\,1.5\,Mpc at $z\sim0.5$). We also excluded potential matches whose cluster redshift was outside the redshift range of our overdensities (with a small padding since clusters just outside our redshift slices would likely influence the density map through the scatter of their photometric redshifts). There were 53 such matches  Table\,\ref{tab:cluster_candidates}; however, several known clusters were matches to more than one of our overdensities due to the generous redshift slices. All cases where a matched cluster was just outside the overdensity redshift slice also had the same cluster matched to another overdensity where it is inside the redshift slice. We chose to keep these double matches since, as described above, massive clusters would influence neighbouring slices due to photometric redshift scatter. 

All together there were 43 matches to unique known clusters. These represent 61\% of the X-ray clusters in the field. As shown in the right-hand panel of Figure\,\ref{fig:cluster_recoverability}, most of the unmatched clusters are at lower redshifts ($z<0.5$) consistent with the expected lower halo masses therein (the XLSSC catalog of \citet{Adami2018} has a limiting cluster mass of log$(M_{\rm halo}/M_{\odot})>13.3$, below our halo mass threshold, but this threshold increases with redshift consistent with the increased recoverability in our density maps. For example, for the smaller sub-set of X-ray clusters in \citet{ClercEtAl2014} (that are all folded into the XLSSC catalog), we recovered 17/21 in our density maps, but found that the ``missing'' four clusters were all associated with density percentiles $>88\%$ -- i.e.\ still significant though below our threshold for finding halos of log$(M_{halo}/M_{\odot})>13.7$. The clusters that are also in \citet{ClercEtAl2014} are marked in Table\,\ref{tab:cluster_candidates}. In addition, our Overdensity \#282 corresponds to XLSS J022303.0−043622 at $z\sim1.22$ first found by \citet{BremerEtAl2006}.  Lastly, \citet{2007MNRAS.382..971V} found a spectroscopic cluster at $z=1.454$ associated with our Overdensity \#311 (more on this structure below). 

\subsubsection{Case studies}
\label{sec:case_studies}
\begin{figure*}
\centering
\begin{minipage}[b]{0.49\textwidth}
\centering
\includegraphics[width=\textwidth]{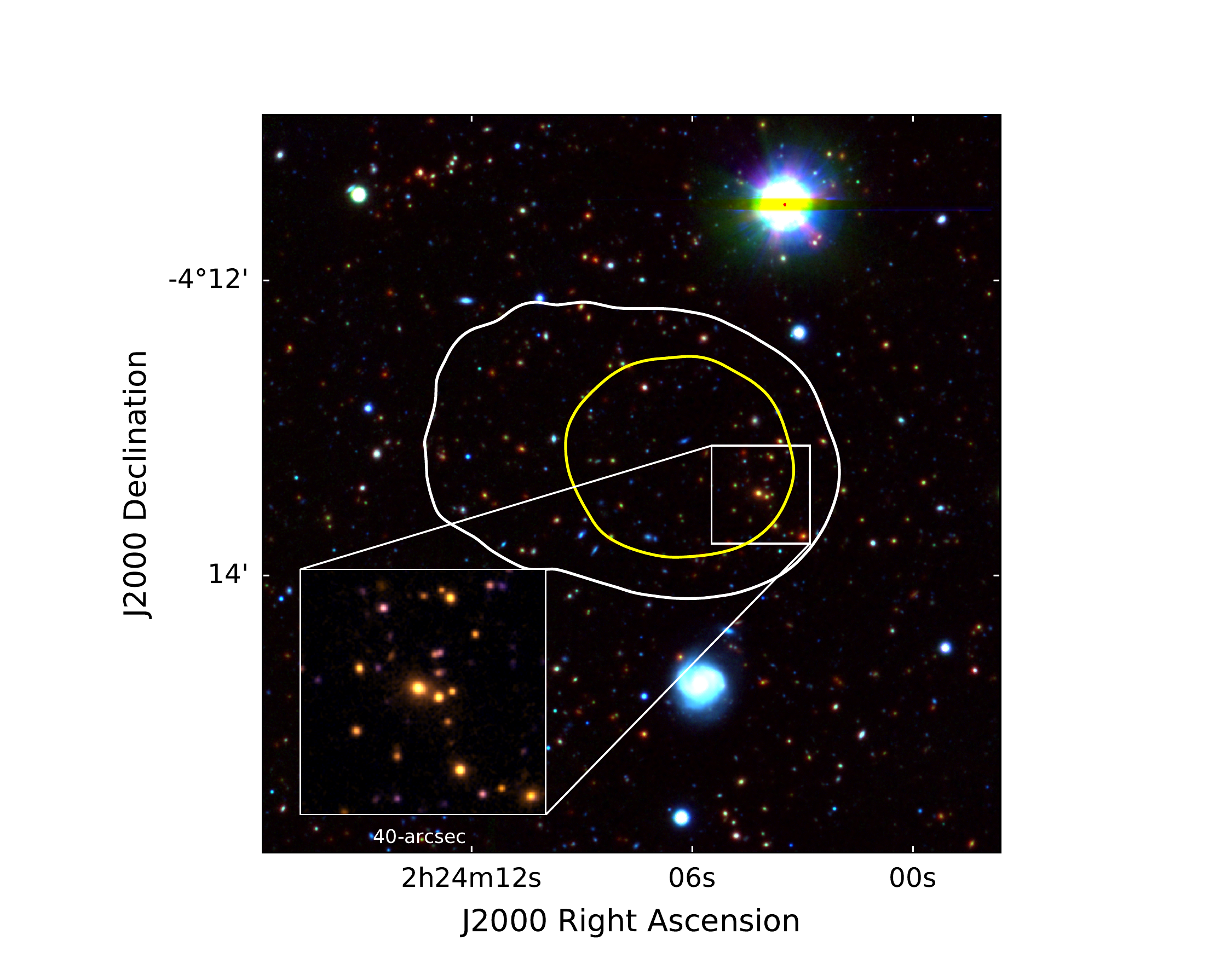}
\end{minipage}
\hfill
\begin{minipage}[b]{0.49\textwidth}
\centering
\includegraphics[width=\textwidth]{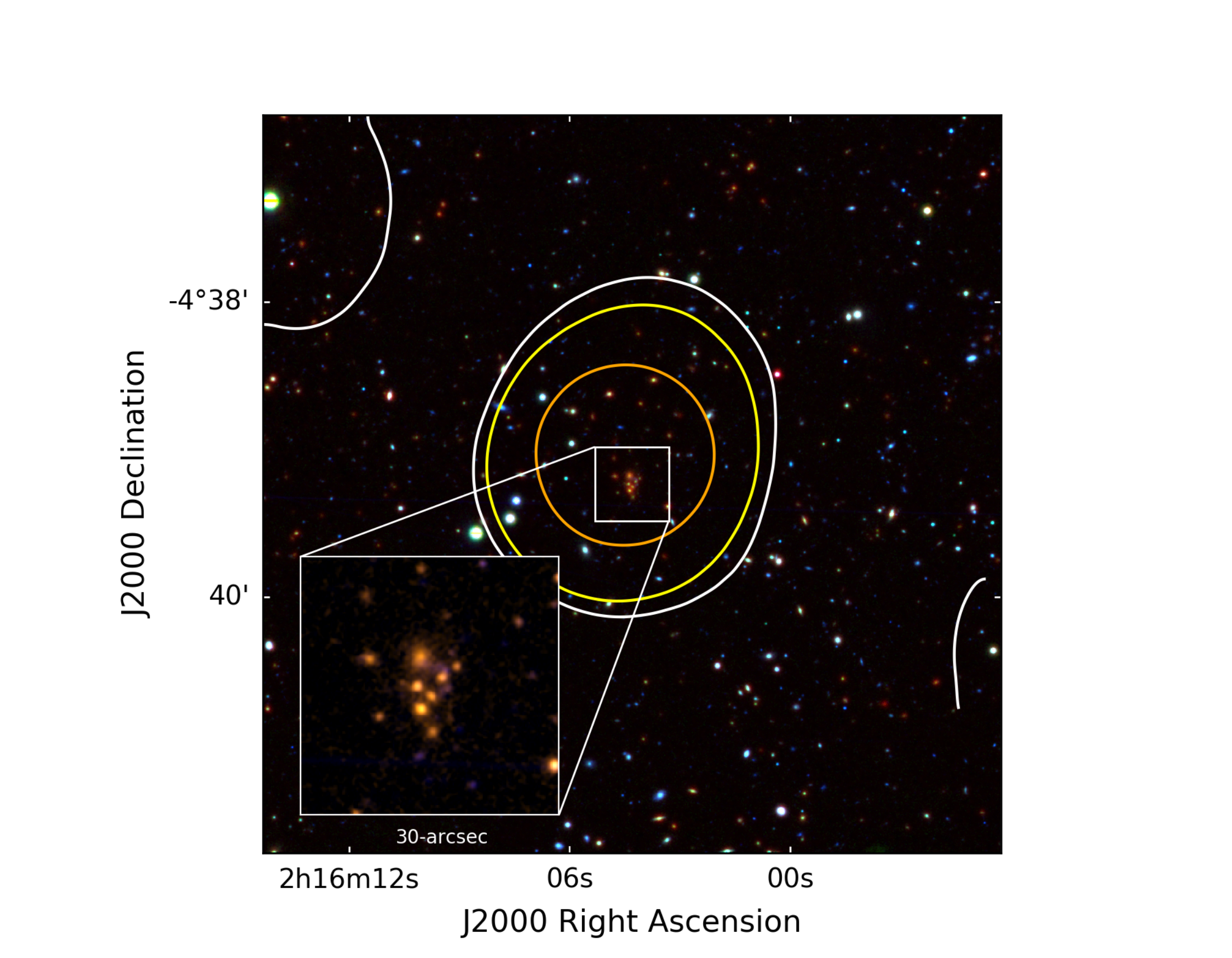}
\end{minipage}
\caption{Two distant galaxy clusters corresponding to overdensities in the XMM-LSS field. {\em Left} the field of the 
$z=1.05$ X-ray detected cluster XLSSC 029 of \citet{ClercEtAl2014}, with contours of overdensity 0.6 in white and 0.8 in yellow.
{\em Right} the highest overdensity in the $1.372<z<1.564$ redshift bin, with contours of overdensity as follows; 0.25 in white, 0.5 in yellow and 1.0 in orange. In both cases, the red channel of the RGB image is {\em Spitzer} IRAC 4.5~$\mu$m, the green VIDEO $H$-band and the blue HSC ultradeep $i$-band. The insets show zoom-ins of the compact 
cluster cores, with VIDEO $K_s$ data in red, VIDEO $H$-band data in green and HSC ultradeep $i$-band data in blue.}
\label{fig:cluster336}
\end{figure*}

Detailed discussion of the clusters and large-scale structures is reserved for a subsequent paper in this series. However, as an illustration of the potential of our technique to find distant clusters, in Figure \ref{fig:cluster336} we show the overdensity corresponding to the highest redshift X-ray detected cluster in the field from \citet{ClercEtAl2014}, XLSSC 029 at $z=1.05$, and an example of what appears to be a compact cluster core associated with the highest percentile overdensity in our highest redshift bin ($1.372<z<1.564$, Overdensity \#336). 

As discussed in Section\,\ref{sec:newClusters}, a quarter of our overdensities show extended morphologies which are likely the result of projection effects -- i.e. 3D structures that are unassociated though overlap on the sky, but can also include interested potentially associated structures such as cluster mergers. We examine one of our extended sources in more detail (Overdensity \#300) which is chosen because it is among our higher redshift structures but also has significant spectroscopic redshift coverage. It allows us an example of how by looking at the structures in overlapping redshift slices we get a coarse view of the 3D structure. 

\begin{figure*}[htb!]{}
\centering
  \includegraphics[width=0.95\textwidth]{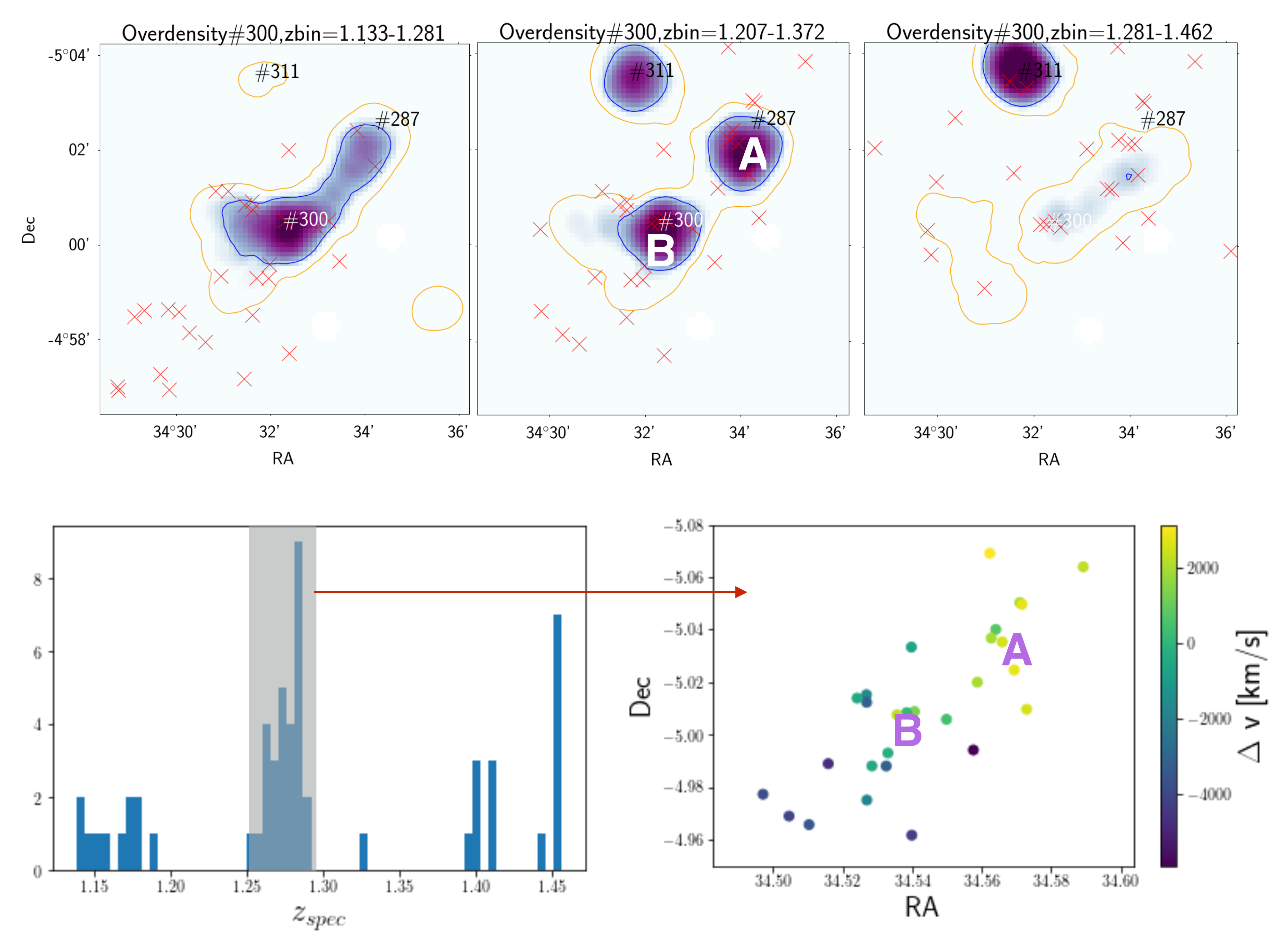}
  \caption{A potential new $z=1.28$ cluster merger. The middle panel shows the peak density redshift slice for this structure, with the neighboring overlapping redshift slices shown on either side. The bottom-left shows all available spectroscopic redshifts showing a clear peak at $z\sim1.28$, corresponding to the middle of the peak density redshift slice. The bottom-right panel shows that Component A is behind Component B in redshift space. 
  \label{fig:z128_merger}}
\end{figure*}

\begin{figure}
%\centering
\includegraphics[angle=-90,scale=0.8,trim=2cm 2cm 1.5cm 2cm, clip=true]{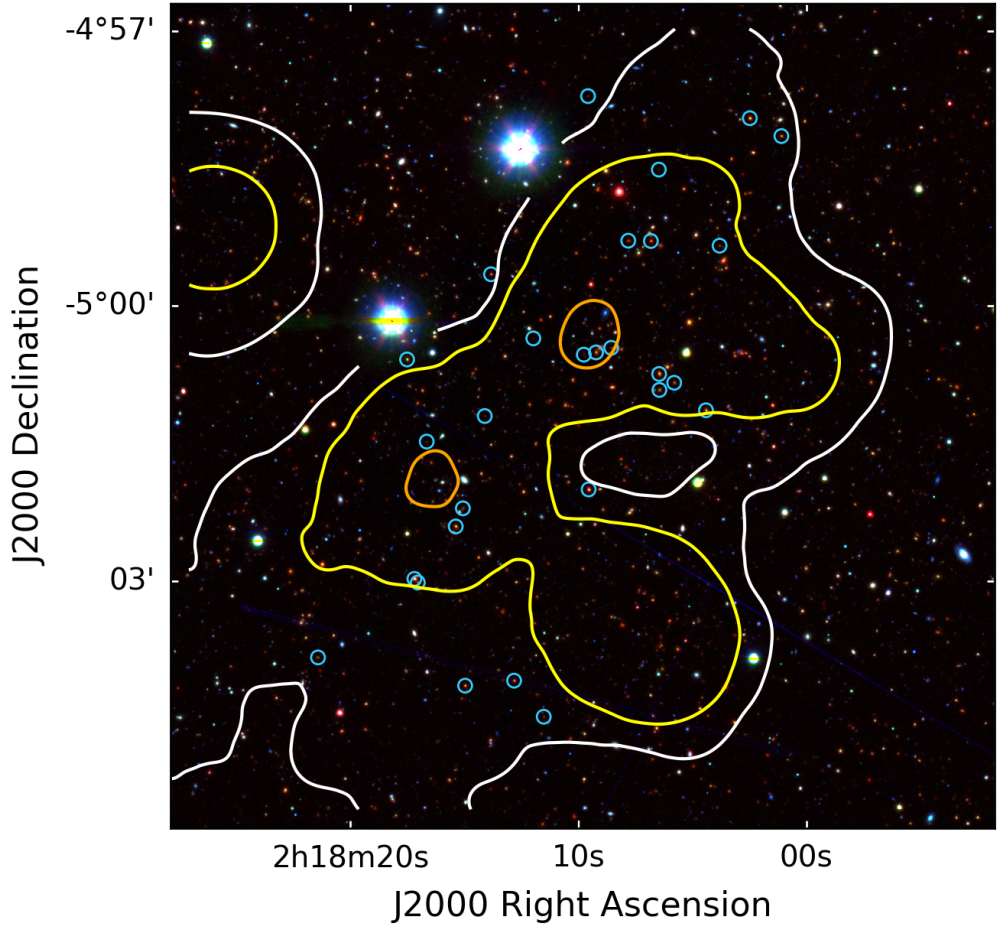}
  \caption{A three color RGB image of the field of the $z=1.28$ cluster merger, with contours of overdensity in the $1.207<z<1.372$ redshift slice as follows; 0.25 in white, 0.5 in yellow and 1.0 in orange. The red channel is {\em Spitzer} IRAC 4.5~$\mu$m, the green VIDEO $H$-band and the blue HSC ultradeep $i$-band. Galaxies with spectroscopic $z\approx 1.28$ are marked with light blue circles.}
  \label{fig:merger_color}
\end{figure}

Examining this overdensity more closely reveals a potential cluster merger at $z=1.28$. This is shown in Figures\,\ref{fig:z128_merger} and \ref{fig:merger_color}).
In Figure\,\ref{fig:z128_merger} we show this structure in three overlapping redshift slices shown on either side. The field had been previously noted as overdense by \citet{2007MNRAS.382..971V}, who obtained spectroscopy of galaxies in the field with the Keck Telescope. We find that the bi-polar overdensity structure in our maps corresponds to a peak in the distribution of spectroscopic redshifts at $z\sim1.28$ (bottom-left), and Figure \ref{fig:merger_color} shows that the distribution of objects at this redshift on the sky corresponds well to the overdensity map.
The relative line of sight velocity (relative to the median in this peak) is shown on a RA-DEC plot highlighting the gradient across this structure. Component A has a median spectroscopic redshift of 1.286 with a velocity dispersion of 670\,km\,s$^{-1}$. Component B has a median redshift of 1.276 with a velocity dispersion of $\sim$\,1100\,km\,s$^{-1}$ (likely affected by the overlapping structure at lower redshift clearly visible in the lower redshift slice). Even without correcting for this structure in its median redshift (since we cannot cleanly disentangle these structures), the relative velocity between these two components is 3000~km\,s$^{-1}$. To check if this number is reasonable for a potential cluster merger, we compared it with the relative velocity of the `bullet' and the larger cluster in the well-studied Bullet Cluster. This is estimated at 2700\,km\,s$^{-1}$ \citep{Springel2007_bullet}. This is the true relative velocity rather than the line of sight one as in our case, but it does show that our number is of reasonable magnitude, especially considering that the median redshift of Component B is likely pulled down by the slightly foreground structure. 

 Figure\,\ref{fig:z128_merger} also helps illustrate how structures at slightly higher and lower redshifts `bleed' into any given redshift slice, but the overlapping slices help us discern some of the 3D structure. For example, Overdensity \#311 sits on top of Components A and B in the middle maps. It is however increasingly stronger in the two overlapping density slices centered at higher redshifts, suggesting what we see here is coming in from a strong overdensity that is actually more distant. Indeed, \citet{2007MNRAS.382..971V} find six objects in the field at $z\approx 1.45$ confirming that Overdensity \#311 is background to our potential merger.

This particular structure is chosen due to the quality of spectroscopic coverage which makes it more likely to be a true physical association. There are other cluster merger candidates among our potential cluster catalog; however, many or most of them are likely line of sight projections (given the wide redshift slices). The XMM-LSS field has ongoing further spectroscopic coverage from the DEVILS survey \citep{Davies2018} and will have even more extensive coverage from the planned Prime Focus Spectrograph \citep[PFS;][]{Tamura2018} galaxy evolution survey. These will help us further disentangle the nature of the overdensity we find our density maps.  

\footnotesize
\begin{TableNotes}
    \item [a] The full redshift range within which overdensity is found.
    \item [b] These are only high quality spectroscopic redshifts within a circle of 750~kpc (proper) of the overdensity center.
    \item [c] Here we reference: \citet[][; here A18]{Adami2018}, \citet[][;here  C14]{ClercEtAl2014}, \citet[][; here B06]{BremerEtAl2006} and \citet[][; here vB07]{2007MNRAS.382..971V}.
\end{TableNotes}
\begin{longtable}{@{\extracolsep{\fill}}L{0.2cm}C{0.8cm}C{1.1cm}C{2cm}C{0.2cm}C{0.8cm}C{4.0cm}}
\caption{Overdensities selected in the XMM-LSS field}\\
\hline
\hline
ID & RA & Dec & $z$-range\tablenotemark{a} & $N_{spec}$\tablenotemark{b} & morphology & References\tablenotemark{c}\\
\hline
\endfirsthead
\multicolumn{7}{c}%
{\tablename\ \thetable\ -- \textit{Continued from previous page}} \\
\hline
\hline
ID & RA & Dec & $z$-range\tablenotemark{a} & $N_{spec}$\tablenotemark{b} & morphology & References\tablenotemark{c}\\
\hline
\endhead
\hline \multicolumn{7}{r}{\textit{Continued on next page}} \\
\endfoot
\insertTableNotes
\endlastfoot
1 & 36.7988 & -5.0698 & 0.100-0.196 & 3 &   &   \\
2 & 36.3951 & -4.2480 & 0.100-0.196 & 13 &   & XLSSC41.0,$z=$0.142,[C14,A18] \\
3 & 34.0523 & -4.2236 & 0.100-0.196 & 3 &   & XLSSC57.0,$z=$0.153,[C14,A18] \\
4 & 36.4396 & -4.3284 & 0.123-0.223 & 18 & extended &   \\
5 & 35.3947 & -4.5792 & 0.146-0.251 & 2 & extended & XLSSC119.0,$z=$0.158,[A18] \\
6 & 35.5526 & -4.2187 & 0.146-0.251 & 2 &   &   \\
7 & 35.0743 & -4.6531 & 0.146-0.251 & 3 &   &   \\
8 & 34.8840 & -4.5052 & 0.146-0.251 & 6 &   &   \\
9 & 36.4270 & -4.9823 & 0.171-0.282 & 2 &   &   \\
10 & 35.4948 & -4.4999 & 0.171-0.282 & 0 & extended &   \\
11 & 36.4270 & -4.5782 & 0.171-0.282 & 7 & extended &   \\
12 & 36.4187 & -4.4133 & 0.171-0.282 & 13 & extended &   \\
13 & 35.4575 & -4.2525 & 0.171-0.282 & 1 & extended &   \\
14 & 36.6714 & -4.2154 & 0.171-0.282 & 13 &   &   \\
15 & 34.9851 & -5.4770 & 0.171-0.282 & 4 &   &   \\
16 & 35.2793 & -4.6730 & 0.171-0.282 & 5 & extended &   \\
17 & 34.3554 & -4.6524 & 0.171-0.282 & 1 &   & XLSSC141.0,$z=$0.196,[A18] \\
18 & 34.9727 & -4.4421 & 0.171-0.282 & 1 &   &   \\
19 & 36.4315 & -5.2302 & 0.196-0.313 & 2 &   &   \\
20 & 37.0447 & -4.8333 & 0.196-0.313 & 6 &   & XLSSC27.0,$z=$0.295,[C14,A18] \\
21 & 36.3563 & -4.6723 & 0.196-0.313 & 10 & extended & XLSSC25.0,$z=$0.265,[C14,A18] \\
22 & 35.5211 & -4.6386 & 0.196-0.313 & 0 & extended &   \\
23 & 35.5249 & -4.5675 & 0.196-0.313 & 1 & extended & XLSSC40.0,$z=$0.32,[A18] \\
24 & 36.8642 & -4.5563 & 0.196-0.313 & 21 &   & XLSSC13.0,$z=$0.308,[C14,A18] \\
25 & 36.1456 & -4.1781 & 0.196-0.313 & 6 &   &   \\
26 & 35.9537 & -4.1594 & 0.196-0.313 & 0 &   &   \\
27 & 36.7776 & -4.1070 & 0.196-0.313 & 9 & extended &   \\
28 & 34.0087 & -5.4997 & 0.196-0.313 & 2 &   &   \\
29 & 35.1787 & -4.6536 & 0.196-0.313 & 4 &   &   \\
30 & 36.9131 & -4.8701 & 0.223-0.347 & 2 &   & XLSSC22.0,$z=$0.293,[C14,A18] \\
31 & 35.4098 & -4.4807 & 0.223-0.347 & 1 & extended & XLSSC126.0,$z=$0.29,[A18] \\
32 & 36.1512 & -4.2382 & 0.223-0.347 & 7 &   & XLSSC44.0,$z=$0.263,[C14,A18] \\
33 & 35.2725 & -5.0955 & 0.223-0.347 & 0 &   &   \\
34 & 37.0243 & -5.2979 & 0.251-0.382 & 3 &   & XLSSC121.0,$z=$0.317,[A18] \\
35 & 35.7522 & -4.1362 & 0.251-0.382 & 5 &   & XLSSC24.0,$z=$0.291,[A18] \\
36 & 34.9909 & -5.4684 & 0.251-0.382 & 5 &   &   \\
37 & 34.9338 & -4.8875 & 0.251-0.382 & 18 &   & XLSSC58.0,$z=$0.332,[C14,A18] \\
38 & 35.3403 & -5.4284 & 0.282-0.421 & 0 &   &   \\
39 & 36.0283 & -5.0744 & 0.282-0.421 & 2 &   & XLSSC18.0,$z=$0.324,[C14,A18] \\
40 & 36.2282 & -4.9369 & 0.282-0.421 & 2 &   &   \\
41 & 36.6545 & -4.8842 & 0.282-0.421 & 2 &   &   \\
42 & 37.0249 & -4.8462 & 0.282-0.421 & 4 &   & XLSSC27.0,$z=$0.295,[C14,A18] \\
43 & 35.5138 & -4.5624 & 0.282-0.421 & 3 & extended & XLSSC40.0,$z=$0.32,[C14,A18] \\
44 & 36.8897 & -4.5507 & 0.282-0.421 & 10 & extended & XLSSC13.0,$z=$0.308,[C14,A18] \\
45 & 36.6368 & -4.4951 & 0.282-0.421 & 13 &   &   \\
46 & 34.0261 & -5.0364 & 0.282-0.421 & 5 &   &   \\
47 & 36.6363 & -4.9954 & 0.313-0.460 & 3 &   &   \\
48 & 35.4938 & -4.9653 & 0.313-0.460 & 1 &   &   \\
49 & 36.3775 & -4.3680 & 0.313-0.460 & 7 &   &   \\
50 & 36.7327 & -4.1680 & 0.313-0.460 & 3 & extended & XLSSC33.0,$z=$0.345,[A18] \\
51 & 36.6115 & -4.1242 & 0.313-0.460 & 6 & extended & XLSSC14.0,$z=$0.344,[A18] \\
52 & 34.0016 & -5.3844 & 0.313-0.460 & 2 &   &   \\
53 & 34.8330 & -5.1269 & 0.313-0.460 & 7 &   &   \\
54 & 34.1833 & -5.0255 & 0.313-0.460 & 9 &   &   \\
55 & 34.0401 & -4.7844 & 0.313-0.460 & 7 &   &   \\
56 & 34.2439 & -4.6338 & 0.313-0.460 & 5 &   &   \\
57 & 34.1585 & -4.4529 & 0.313-0.460 & 8 &   & XLSSC144.0,$z=$0.447,[A18] \\
58 & 36.5369 & -5.0739 & 0.347-0.504 & 5 &   &   \\
59 & 36.7694 & -4.9737 & 0.347-0.504 & 2 &   &   \\
60 & 35.9480 & -4.5187 & 0.347-0.504 & 3 &   & XLSSC26.0,$z=$0.435,[A18] \\
61 & 34.9174 & -5.3952 & 0.347-0.504 & 4 &   &   \\
62 & 34.2278 & -4.5521 & 0.347-0.504 & 7 & extended & XLSSC194.0,$z=$0.411,[A18] \\
63 & 34.4138 & -4.5084 & 0.347-0.504 & 3 &   &   \\
64 & 35.6449 & -5.3939 & 0.382-0.549 & 7 &   &   \\
65 & 36.0653 & -5.3793 & 0.382-0.549 & 10 &   & XLSSC19.0,$z=$0.496,[A18] \\
66 & 36.1362 & -4.8345 & 0.382-0.549 & 17 &   & XLSSC53.0,$z=$0.495,[C14,A18] \\
67 & 36.2364 & -4.7469 & 0.382-0.549 & 6 &   &   \\
68 & 36.1019 & -4.7299 & 0.382-0.549 & 11 &   &   \\
69 & 35.9968 & -4.6155 & 0.382-0.549 & 5 & extended & XLSSC49.0,$z=$0.494,[C14,A18] \\
70 & 36.6616 & -4.5207 & 0.382-0.549 & 15 &   &   \\
71 & 36.3537 & -4.3966 & 0.382-0.549 & 5 &   & XLSSC42.0,$z=$0.463,[A18] \\
72 & 34.7479 & -5.4814 & 0.382-0.549 & 10 & extended & XLSSC142.0,$z=$0.451,[A18] \\
73 & 34.3935 & -4.9926 & 0.382-0.549 & 14 &   &   \\
74 & 34.4033 & -4.8174 & 0.382-0.549 & 8 &   & XLSSC124.0,$z=$0.516,[A18] \\
75 & 34.6257 & -4.2458 & 0.382-0.549 & 2 &   &   \\
76 & 34.7234 & -4.2118 & 0.382-0.549 & 4 &   &   \\
77 & 36.0249 & -5.2895 & 0.421-0.599 & 11 &   &   \\
78 & 36.7449 & -5.1190 & 0.421-0.599 & 13 &   &   \\
79 & 36.1013 & -5.0107 & 0.421-0.599 & 6 &   &   \\
80 & 36.6407 & -4.9992 & 0.421-0.599 & 7 &   & XLSSC20.0,$z=$0.494,[A18] \\
81 & 35.3767 & -4.4993 & 0.421-0.599 & 10 &   &   \\
82 & 36.8907 & -4.3611 & 0.421-0.599 & 6 &   &   \\
83 & 36.7565 & -4.1722 & 0.421-0.599 & 11 &   &   \\
84 & 34.1776 & -5.0038 & 0.421-0.599 & 9 &   &   \\
85 & 34.4507 & -4.5270 & 0.421-0.599 & 6 &   &   \\
86 & 34.1614 & -4.4694 & 0.421-0.599 & 9 & extended & XLSSC144.0,$z=$0.447,[A18] \\
87 & 36.3098 & -5.4046 & 0.460-0.649 & 9 &   &   \\
88 & 36.5042 & -5.2969 & 0.460-0.649 & 10 &   &   \\
89 & 36.1662 & -5.0616 & 0.460-0.649 & 12 & extended &   \\
90 & 36.3628 & -4.7143 & 0.460-0.649 & 9 &   &   \\
91 & 35.9608 & -4.6550 & 0.460-0.649 & 6 & extended &   \\
92 & 36.0005 & -4.5824 & 0.460-0.649 & 8 & extended & XLSSC49.0,$z=$0.494,[C14,A18] \\
93 & 36.3805 & -4.2681 & 0.460-0.649 & 11 &   & XLSSC45.0,$z=$0.556,[A18] \\
94 & 34.6818 & -5.0660 & 0.460-0.649 & 22 & extended &   \\
95 & 34.3460 & -5.0353 & 0.460-0.649 & 14 &   &   \\
96 & 34.7193 & -4.9957 & 0.460-0.649 & 12 & extended &   \\
97 & 34.6575 & -4.9891 & 0.460-0.649 & 23 & extended &   \\
98 & 35.0683 & -4.9231 & 0.460-0.649 & 9 &   & XLSSC183.0,$z=$0.511,[A18] \\
99 & 34.4962 & -4.8638 & 0.460-0.649 & 5 &   &   \\
100 & 35.1829 & -5.4278 & 0.504-0.705 & 11 &   & XLSSC130.0,$z=$0.546,[A18] \\
101 & 36.1409 & -4.8335 & 0.504-0.705 & 2 &   & XLSSC53.0,$z=$0.495,[A18] \\
102 & 36.8880 & -4.5457 & 0.504-0.705 & 23 &   &   \\
103 & 36.5588 & -4.2664 & 0.504-0.705 & 19 &   &   \\
104 & 36.1663 & -4.2265 & 0.504-0.705 & 3 &   & XLSSC131.0,$z=$0.513,[A18] \\
105 & 36.9914 & -4.1740 & 0.504-0.705 & 5 &   &   \\
106 & 34.4421 & -5.4362 & 0.504-0.705 & 14 &   &   \\
107 & 34.5962 & -5.2871 & 0.504-0.705 & 13 &   &   \\
108 & 34.9381 & -5.2577 & 0.504-0.705 & 8 &   &   \\
109 & 34.7798 & -5.2472 & 0.504-0.705 & 7 & extended &   \\
110 & 34.6405 & -5.2388 & 0.504-0.705 & 15 &   &   \\
111 & 34.7102 & -5.1422 & 0.504-0.705 & 13 & extended &   \\
112 & 34.2839 & -4.9385 & 0.504-0.705 & 5 & extended &   \\
113 & 34.4780 & -4.9280 & 0.504-0.705 & 8 &   &   \\
114 & 34.2733 & -4.8608 & 0.504-0.705 & 15 &   &   \\
115 & 34.4991 & -4.8062 & 0.504-0.705 & 10 & extended &   \\
116 & 34.3366 & -4.7263 & 0.504-0.705 & 11 &   &   \\
117 & 35.1027 & -4.2580 & 0.504-0.705 & 10 &   &   \\
118 & 36.2049 & -4.9554 & 0.549-0.760 & 8 &   &   \\
119 & 36.3672 & -4.7474 & 0.549-0.760 & 12 &   &   \\
120 & 36.7142 & -4.4910 & 0.549-0.760 & 18 &   &   \\
121 & 35.7909 & -4.2204 & 0.549-0.760 & 7 &   & XLSSC30.0,$z=$0.631,[A18] \\
122 & 34.6100 & -5.4219 & 0.549-0.760 & 13 & extended & XLSSC80.0,$z=$0.646,[C14,A18] \\
123 & 34.3929 & -5.4421 & 0.549-0.760 & 8 & extended &   \\
124 & 34.3056 & -5.4279 & 0.549-0.760 & 5 & extended &   \\
125 & 34.3503 & -5.4178 & 0.549-0.760 & 9 & extended &   \\
126 & 34.2508 & -5.4178 & 0.549-0.760 & 10 & extended &   \\
127 & 34.9407 & -5.3007 & 0.549-0.760 & 12 & extended &   \\
128 & 34.4091 & -5.2321 & 0.549-0.760 & 19 & extended & XLSSC59.0,$z=$0.645,[C14,A18] \\
129 & 34.5532 & -5.2502 & 0.549-0.760 & 12 & extended &   \\
130 & 34.4822 & -5.2159 & 0.549-0.760 & 13 & extended &   \\
131 & 34.3706 & -5.1998 & 0.549-0.760 & 21 & extended & XLSSC59.0,$z=$0.645,[A18] \\
132 & 34.2021 & -5.1634 & 0.549-0.760 & 13 &   &   \\
133 & 34.2772 & -5.0746 & 0.549-0.760 & 9 & extended &   \\
134 & 34.1879 & -5.0463 & 0.549-0.760 & 10 &   &   \\
135 & 34.5024 & -4.9615 & 0.549-0.760 & 4 &   &   \\
136 & 34.0134 & -4.5132 & 0.549-0.760 & 8 &   &   \\
137 & 36.1709 & -4.6442 & 0.599-0.820 & 7 &   &   \\
138 & 36.1103 & -4.3505 & 0.599-0.820 & 6 &   &   \\
139 & 36.2862 & -4.2358 & 0.599-0.820 & 5 &   &   \\
140 & 34.5741 & -5.5214 & 0.599-0.820 & 4 & extended &   \\
141 & 34.2360 & -5.4825 & 0.599-0.820 & 6 &   &   \\
142 & 34.4959 & -5.4650 & 0.599-0.820 & 13 & extended &   \\
143 & 34.4256 & -5.4708 & 0.599-0.820 & 10 & extended &   \\
144 & 34.7168 & -5.4572 & 0.599-0.820 & 5 &   &   \\
145 & 34.1383 & -5.4280 & 0.599-0.820 & 12 &   &   \\
146 & 34.5370 & -5.3561 & 0.599-0.820 & 4 & extended &   \\
147 & 34.6308 & -5.3386 & 0.599-0.820 & 6 & extended &   \\
148 & 35.0686 & -5.1810 & 0.599-0.820 & 4 & extended &   \\
149 & 34.2946 & -5.1441 & 0.599-0.820 & 9 & extended &   \\
150 & 34.6445 & -5.0215 & 0.599-0.820 & 19 & extended &   \\
151 & 34.6464 & -4.9632 & 0.599-0.820 & 8 & extended &   \\
152 & 34.0562 & -4.8387 & 0.599-0.820 & 2 &   &   \\
153 & 34.0347 & -4.7940 & 0.599-0.820 & 14 &   &   \\
154 & 34.2653 & -4.4769 & 0.599-0.820 & 10 &   & XLSSC195.0,$z=$0.661,[A18] \\
155 & 36.9854 & -5.3820 & 0.649-0.879 & 15 &   &   \\
156 & 36.2831 & -5.2991 & 0.649-0.879 & 4 &   &   \\
157 & 36.9533 & -4.9092 & 0.649-0.879 & 5 &   &   \\
158 & 35.4010 & -4.7679 & 0.649-0.879 & 6 &   &   \\
159 & 35.3896 & -4.5983 & 0.649-0.879 & 3 &   &   \\
160 & 36.5557 & -4.4212 & 0.649-0.879 & 5 &   &   \\
161 & 36.1790 & -4.2084 & 0.649-0.879 & 4 &   &   \\
162 & 36.2945 & -4.1424 & 0.649-0.879 & 1 &   &   \\
163 & 34.5169 & -5.5214 & 0.649-0.879 & 3 & extended &   \\
164 & 34.7517 & -5.1409 & 0.649-0.879 & 8 &   &   \\
165 & 35.2230 & -5.0768 & 0.649-0.879 & 5 &   &   \\
166 & 34.6078 & -4.9845 & 0.649-0.879 & 16 & extended & XLSSC64.0,$z=$0.874,[A18] \\
167 & 34.4128 & -4.5795 & 0.649-0.879 & 9 &   &   \\
168 & 34.0039 & -4.5738 & 0.649-0.879 & 6 &   &   \\
169 & 34.8804 & -4.4457 & 0.649-0.879 & 6 &   &   \\
170 & 34.8785 & -4.3101 & 0.649-0.879 & 2 &   &   \\
171 & 34.0834 & -4.2574 & 0.649-0.879 & 3 &   &   \\
172 & 36.9232 & -5.2798 & 0.705-0.942 & 3 &   &   \\
173 & 35.9436 & -5.0201 & 0.705-0.942 & 7 & extended & XLSSC15.0,$z=$0.858,[A18] \\
174 & 35.6752 & -5.0165 & 0.705-0.942 & 4 & extended & XLSSC71.0,$z=$0.833,[A18] \\
175 & 35.6477 & -4.9689 & 0.705-0.942 & 4 & extended & XLSSC71.0,$z=$0.833,[A18] \\
176 & 35.8149 & -4.9561 & 0.705-0.942 & 5 &   &   \\
177 & 35.6219 & -4.7183 & 0.705-0.942 & 1 &   &   \\
178 & 36.5501 & -4.4495 & 0.705-0.942 & 6 &   &   \\
179 & 35.3224 & -4.4001 & 0.705-0.942 & 5 & extended &   \\
180 & 34.1810 & -5.3731 & 0.705-0.942 & 9 &   &   \\
181 & 34.2178 & -5.2049 & 0.705-0.942 & 12 &   &   \\
182 & 34.8408 & -5.0805 & 0.705-0.942 & 13 &   &   \\
183 & 35.2250 & -5.0238 & 0.705-0.942 & 5 & extended &   \\
184 & 35.2268 & -4.9049 & 0.705-0.942 & 2 & extended &   \\
185 & 35.1147 & -4.8628 & 0.705-0.942 & 4 & extended &   \\
186 & 35.2139 & -4.8464 & 0.705-0.942 & 5 & extended &   \\
187 & 34.9291 & -4.8391 & 0.705-0.942 & 11 &   &   \\
188 & 35.0742 & -4.8189 & 0.705-0.942 & 3 & extended &   \\
189 & 35.0283 & -4.8189 & 0.705-0.942 & 4 & extended &   \\
190 & 34.7747 & -4.7092 & 0.705-0.942 & 9 &   &   \\
191 & 35.2305 & -4.5848 & 0.705-0.942 & 1 &   &   \\
192 & 34.8445 & -4.5281 & 0.705-0.942 & 7 &   &   \\
193 & 34.8261 & -4.4934 & 0.705-0.942 & 8 & extended &   \\
194 & 35.3187 & -4.4458 & 0.705-0.942 & 2 & extended &   \\
195 & 35.1992 & -4.3105 & 0.705-0.942 & 0 &   &   \\
196 & 34.0211 & -4.2739 & 0.705-0.942 & 0 &   &   \\
197 & 34.3023 & -4.2135 & 0.705-0.942 & 10 &   &   \\
198 & 35.4056 & -5.3048 & 0.760-1.004 & 3 & extended &   \\
199 & 35.9882 & -5.1336 & 0.760-1.004 & 4 &   &   \\
200 & 35.9255 & -5.1104 & 0.760-1.004 & 6 &   &   \\
201 & 35.9721 & -4.9748 & 0.760-1.004 & 1 & extended &   \\
202 & 36.2894 & -4.8000 & 0.760-1.004 & 2 &   &   \\
203 & 36.0312 & -4.2559 & 0.760-1.004 & 0 &   &   \\
204 & 35.3267 & -4.2434 & 0.760-1.004 & 10 &   & XLSSC184.0,$z=$0.811,[A18] \\
205 & 34.4484 & -5.5082 & 0.760-1.004 & 6 &   &   \\
206 & 35.2156 & -5.1728 & 0.760-1.004 & 4 &   &   \\
207 & 34.9539 & -5.1389 & 0.760-1.004 & 13 &   &   \\
208 & 35.3070 & -5.0979 & 0.760-1.004 & 1 &   &   \\
209 & 34.6366 & -5.0033 & 0.760-1.004 & 26 &   & XLSSC64.0,$z=$0.874,[A18] \\
210 & 35.3375 & -4.9266 & 0.760-1.004 & 2 &   &   \\
211 & 35.0722 & -4.8963 & 0.760-1.004 & 8 & extended &   \\
212 & 35.0453 & -4.8785 & 0.760-1.004 & 7 & extended &   \\
213 & 34.9557 & -4.8232 & 0.760-1.004 & 7 &   &   \\
214 & 34.5183 & -4.7250 & 0.760-1.004 & 1 &   &   \\
215 & 35.1547 & -4.5217 & 0.760-1.004 & 7 &   &   \\
216 & 34.6384 & -4.4735 & 0.760-1.004 & 1 &   &   \\
217 & 35.2102 & -4.4307 & 0.760-1.004 & 2 & extended &   \\
218 & 34.3426 & -4.4039 & 0.760-1.004 & 4 &   &   \\
219 & 35.2246 & -4.3754 & 0.760-1.004 & 4 & extended &   \\
220 & 36.9468 & -5.3541 & 0.820-1.069 & 0 &   &   \\
221 & 36.6085 & -4.7522 & 0.820-1.069 & 8 &   &   \\
222 & 36.3876 & -4.4259 & 0.820-1.069 & 4 & extended &   \\
223 & 36.4349 & -4.3858 & 0.820-1.069 & 4 & extended &   \\
224 & 35.2217 & -5.2250 & 0.820-1.069 & 1 & extended &   \\
225 & 35.2498 & -5.2198 & 0.820-1.069 & 0 & extended &   \\
226 & 34.3592 & -5.2041 & 0.820-1.069 & 13 &   &   \\
227 & 34.2733 & -5.1674 & 0.820-1.069 & 8 &   &   \\
228 & 34.8097 & -4.5830 & 0.820-1.069 & 5 &   &   \\
229 & 35.0744 & -4.4190 & 0.820-1.069 & 3 &   &   \\
230 & 34.1961 & -4.3963 & 0.820-1.069 & 0 &   &   \\
231 & 35.2628 & -5.2960 & 0.879-1.133 & 0 & extended &   \\
232 & 36.8115 & -5.2704 & 0.879-1.133 & 1 &   &   \\
233 & 37.0215 & -5.0923 & 0.879-1.133 & 1 &   &   \\
234 & 36.3366 & -4.2737 & 0.879-1.133 & 0 &   &   \\
235 & 35.2043 & -4.2035 & 0.879-1.133 & 0 &   &   \\
236 & 36.8425 & -4.1829 & 0.879-1.133 & 0 &   &   \\
237 & 35.0666 & -5.2755 & 0.879-1.133 & 1 &   &   \\
238 & 35.1681 & -5.2087 & 0.879-1.133 & 1 & extended &   \\
239 & 35.2232 & -5.0580 & 0.879-1.133 & 1 &   &   \\
240 & 34.8687 & -4.8525 & 0.879-1.133 & 4 &   &   \\
241 & 35.0459 & -4.7343 & 0.879-1.133 & 3 &   &   \\
242 & 35.0356 & -4.6949 & 0.879-1.133 & 1 &   &   \\
243 & 34.1167 & -4.5990 & 0.879-1.133 & 0 &   &   \\
244 & 36.4925 & -4.4876 & 0.942-1.207 & 0 &   &   \\
245 & 36.8058 & -4.3039 & 0.942-1.207 & 1 &   & XLSSC5.0,$z=$1.058,[C14,A18] \\
246 & 36.1420 & -4.2348 & 0.942-1.207 & 1 & extended &   \\
247 & 35.5831 & -4.2668 & 0.942-1.207 & 0 &   &   \\
248 & 36.2808 & -4.2584 & 0.942-1.207 & 0 & extended &   \\
249 & 36.0319 & -4.2230 & 0.942-1.207 & 1 & extended & XLSSC29.0,$z=$1.05,[C14,A18] \\
250 & 36.2605 & -4.1927 & 0.942-1.207 & 0 &   &   \\
251 & 35.2123 & -5.1937 & 0.942-1.207 & 2 &   &   \\
252 & 34.5417 & -5.0521 & 0.942-1.207 & 12 & extended &   \\
253 & 34.6382 & -5.0336 & 0.942-1.207 & 13 & extended &   \\
254 & 34.5264 & -5.0184 & 0.942-1.207 & 13 & extended &   \\
255 & 34.5603 & -5.0117 & 0.942-1.207 & 6 & extended &   \\
256 & 34.5958 & -4.9797 & 0.942-1.207 & 5 & extended &   \\
257 & 34.2843 & -4.9611 & 0.942-1.207 & 3 &   &   \\
258 & 34.5874 & -4.8465 & 0.942-1.207 & 5 &   &   \\
259 & 35.1107 & -4.8078 & 0.942-1.207 & 1 & extended &   \\
260 & 34.8770 & -4.7859 & 0.942-1.207 & 2 &   &   \\
261 & 34.4841 & -4.6190 & 0.942-1.207 & 0 &   &   \\
262 & 34.0472 & -4.5365 & 0.942-1.207 & 0 &   &   \\
263 & 35.1564 & -4.5196 & 0.942-1.207 & 2 &   &   \\
264 & 34.1183 & -4.3966 & 0.942-1.207 & 0 &   &   \\
265 & 34.7618 & -4.2028 & 0.942-1.207 & 0 &   &   \\
266 & 36.8379 & -4.6075 & 1.004-1.281 & 1 &   &   \\
267 & 36.4719 & -4.2865 & 1.004-1.281 & 1 &   &   \\
268 & 35.9438 & -4.1668 & 1.004-1.281 & 0 &   &   \\
269 & 34.4881 & -5.1015 & 1.004-1.281 & 4 &   &   \\
270 & 34.1522 & -4.8769 & 1.004-1.281 & 1 &   &   \\
271 & 34.2909 & -4.8353 & 1.004-1.281 & 6 &   &   \\
272 & 34.3009 & -4.7672 & 1.004-1.281 & 2 &   &   \\
273 & 34.5600 & -4.7206 & 1.004-1.281 & 0 &   &   \\
274 & 35.1332 & -4.6491 & 1.004-1.281 & 1 &   &   \\
275 & 34.4179 & -4.6341 & 1.004-1.281 & 0 &   &   \\
276 & 34.6352 & -4.5892 & 1.004-1.281 & 0 &   &   \\
277 & 34.3310 & -4.5060 & 1.004-1.281 & 0 &   &   \\
278 & 36.9064 & -5.2820 & 1.069-1.372 & 0 &   &   \\
279 & 36.5711 & -4.8463 & 1.069-1.372 & 0 &   &   \\
280 & 36.6850 & -4.7016 & 1.069-1.372 & 0 &   &   \\
281 & 35.6113 & -4.6737 & 1.069-1.372 & 0 &   &   \\
282 & 35.7699 & -4.6046 & 1.069-1.372 & 0 &   & XLSSC46.0,$z=$1.217,[B06,A18] \\
283 & 36.8965 & -4.6063 & 1.069-1.372 & 0 &   &   \\
284 & 36.7362 & -4.4008 & 1.069-1.372 & 1 &   &   \\
285 & 36.7759 & -4.3071 & 1.069-1.372 & 0 &   & XLSSC5.0,$z=$1.058,[A18] \\
286 & 36.3266 & -4.1822 & 1.069-1.372 & 0 &   &   \\
287 & 34.5705 & -5.0419 & 1.069-1.372 & 7 & extended &   \\
288 & 34.4549 & -5.0189 & 1.069-1.372 & 3 &   & XLSSC203.0,$z=$1.077,[A18] \\
289 & 34.2633 & -4.8184 & 1.069-1.372 & 5 & extended &   \\
290 & 35.3552 & -4.6967 & 1.069-1.372 & 1 &   &   \\
291 & 35.2759 & -4.6622 & 1.069-1.372 & 0 &   &   \\
292 & 35.0364 & -4.3926 & 1.069-1.372 & 0 &   &   \\
293 & 36.8365 & -5.3993 & 1.133-1.462 & 1 &   &   \\
294 & 36.8512 & -5.1403 & 1.133-1.462 & 0 &   &   \\
295 & 36.7792 & -5.1354 & 1.133-1.462 & 0 &   &   \\
296 & 36.5205 & -4.8992 & 1.133-1.462 & 0 &   &   \\
297 & 35.3616 & -4.4512 & 1.133-1.462 & 1 &   &   \\
298 & 34.9115 & -5.3520 & 1.133-1.462 & 3 &   &   \\
299 & 34.9295 & -5.1419 & 1.133-1.462 & 0 &   &   \\
300 & 34.5382 & -5.0067 & 1.133-1.462 & 14 & extended & pot.merger  \\
    &         &         &             &   &           & $z\approx1.28$\\
301 & 34.5039 & -4.8878 & 1.133-1.462 & 4 &   &   \\
302 & 34.6987 & -4.7102 & 1.133-1.462 & 0 &   &   \\
303 & 34.5923 & -4.6630 & 1.133-1.462 & 2 & extended &   \\
304 & 34.5824 & -4.6353 & 1.133-1.462 & 0 & extended &   \\
305 & 35.1586 & -4.6304 & 1.133-1.462 & 0 &   &   \\
306 & 35.0244 & -4.6076 & 1.133-1.462 & 0 &   &   \\
307 & 35.3387 & -4.5554 & 1.133-1.462 & 0 &   &   \\
308 & 34.5988 & -4.5424 & 1.133-1.462 & 1 &   &   \\
309 & 36.8256 & -5.2332 & 1.207-1.564 & 0 &   &   \\
310 & 34.9422 & -5.5224 & 1.207-1.564 & 0 &   &   \\
311 & 34.5282 & -5.0587 & 1.207-1.564 & 1 & extended & [vB07]  \\
312 & 34.7084 & -5.0038 & 1.207-1.564 & 5 &   &   \\
313 & 34.8480 & -4.8713 & 1.207-1.564 & 2 &   &   \\
314 & 34.6532 & -4.6661 & 1.207-1.564 & 0 & extended &   \\
315 & 35.0120 & -4.6564 & 1.207-1.564 & 0 &   &   \\
316 & 35.0153 & -4.5287 & 1.207-1.564 & 0 &   &   \\
317 & 35.0607 & -4.5158 & 1.207-1.564 & 0 &   &   \\
318 & 34.7409 & -4.4528 & 1.207-1.564 & 0 &   &   \\
319 & 34.8562 & -4.4431 & 1.207-1.564 & 0 &   &   \\
320 & 36.5928 & -4.9511 & 1.281-1.665 & 0 &   &   \\
321 & 36.1908 & -4.9302 & 1.281-1.665 & 1 &   &   \\
322 & 34.8249 & -5.2773 & 1.281-1.665 & 0 &   &   \\
323 & 34.3987 & -5.1761 & 1.281-1.665 & 2 &   &   \\
324 & 34.4988 & -5.1262 & 1.281-1.665 & 2 &   &   \\
325 & 34.2696 & -4.9463 & 1.281-1.665 & 3 &   &   \\
326 & 34.7846 & -4.8290 & 1.281-1.665 & 2 &   &   \\
327 & 34.1937 & -4.7985 & 1.281-1.665 & 1 &   &   \\
328 & 34.5747 & -4.7262 & 1.281-1.665 & 0 &   &   \\
329 & 34.8266 & -4.4498 & 1.281-1.665 & 0 &   &   \\
330 & 34.6425 & -4.3085 & 1.281-1.665 & 0 &   &   \\
331 & 34.7457 & -5.3043 & 1.372-1.665 & 1 &   &   \\
332 & 34.2198 & -5.1827 & 1.372-1.665 & 3 &   &   \\
333 & 34.8148 & -5.0627 & 1.372-1.665 & 3 &   &   \\
334 & 34.5479 & -5.0115 & 1.372-1.665 & 10 &   &   \\
335 & 35.0802 & -4.6834 & 1.372-1.665 & 1 &   &   \\
336 & 34.0172 & -4.6514 & 1.372-1.665 & 1 &   &   \\
337 & 34.5543 & -4.4866 & 1.372-1.665 & 0 &   &   \\
338 & 34.5945 & -4.4674 & 1.372-1.665 & 0 &   &   \\
339 & 34.6428 & -4.4466 & 1.372-1.665 & 0 &   &   \\
\\
\hline 
\hline
\label{tab:cluster_candidates}
\end{longtable}

\normalsize

\section{Summary and conclusions \label{sec:conclusions}}

In this paper we present new photometric redshifts for the 4.8\,deg$^2$ XMM-LSS field. We use them to compute surface density maps in the range $0.1<z<1.6$.  We use a simulated lightcone to assess the recoverability of true structures using such density maps. We summarize our key results as follows:

\begin{enumerate}
  \item Photometric redshifts at the level of $\sigma(z)/(1+z)\approx0.03$ allow us to recover dark matter halos with log$(M_{\rm halo}/M_{\odot})\gtrsim$13.7, as we show based on a comparison with a simulated lightcone. 
  \item We construct 2D density maps for the XMM-LSS in 28 redshift slices covering $z=0.1-1.6$. These density maps show evidence of extended overdensities, visually similar to filaments, as well as compact overdensities likely associated with massive halos.
  \item Using an evolving percentile mass density per comoving volume threshold we determine from our simulated lightcone, we find 339 halos with log$(M_{\rm halo}/M_{\odot})>$13.7 from $0.1 < z < 1.6$ and a peak of $z\sim0.8$. Their number and redshift distribution are consistent with expectations from our simulated lightcone. 
  \item Among our likely massive halo overdensities, we recover 43 of the 70 known spectroscopically-confirmed X-ray clusters in the field \citep{Adami2018}. The unrecovered ones are predominantly below $z\sim0.4$ where the X-ray cluster tend to have lower masses than our target threshold.  
  \item We present some interesting case studies including a potential massive evolved cluster at $z\sim1.5$ as well as as a potential cluster merger at $z\sim1.28$.  
\end{enumerate}

This paper is proof of concept on the degree to which we can reliably probe both the local and large scale environment of galaxies using photometric redshifts of the quality already achievable for moderately large area surveys such as SERVS. The next papers in the series will use these density maps to further quantify the presence of filaments and look at the role of local and large scale environment on the growth of galaxies and their supermassive black holes. 

\acknowledgements
The authors acknowledge useful discussions with Rachel Bezanson, Frank van den Bosch and Eric Gawiser which helped guide our analysis. The authors especially thank Ryan Cybulski for providing his Voronoi Tesselation {\sc idl} code which NK translated into python and has used for this project.  B.D. acknowledges financial support from the National Science Foundation, grant number 1716907.

\appendix
\section{Photometric redshifts comparison}

In this paper we use photometric redshifts based on the template fitting code EAZY as described in detail in the body of the paper. However, this may lead to a potential bias driven by the particular choice of templates. Our zero-point offset corrections are meant to mitigate for such biases to some extent, but this correction is driven by the available spectroscopic redshifts which represent a biased subset of the whole as shown in Figure\,\ref{fig:photoz}. The HELP team performed a more sophisticated photometric redshift analysis in this field considering several different template libraries and performing a hierarchical Bayesian analysis to find the best redshift overall. While this is clearly a more sophisticated approach, the downside for us is that that work uses the data fusion catalogs which adopt the single-band SExtractor IRAC photometry. However, we now have the Tractor forced photometry \citep{NylandEtAl2017} which is better at de-blending IRAC sources leading to more accurate results especially at higher redshifts ($z>1$). 

\begin{figure}
\centering
\includegraphics[scale=0.8,clip=true]{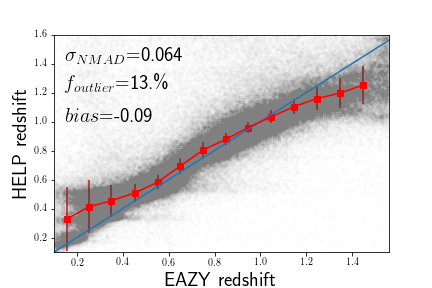}
  \caption{A comparison between the EAZY photometric redshifts adopted in this paper, and the HELP photometric redshifts described in \citet{DuncanEtAl2018}. The legend gives the statistics across the full range of interest ($0.1<z<1.6$), whereas the red symbols and errorbars give the median values and $\sigma_{\rm NMAD}$ in bins of 0.1 in EAZY redshifts. 
  \label{fig:eazy_help}}
\end{figure}

In Figure\,\ref{fig:eazy_help}, we show the direct comparison between our EAZY photometric redshifts and the HELP photometric redshifts\footnote{We adopted the $z1_{\rm median}$ redshifts from the HELP catalog, which is their recommended measure.}. This plot shows very good agreement in the range $0.3<z<1.0$, but also has significant deviations at either end outside this range. We performed the pair analysis for the HELP redshifts at those two ends -- for the $z<0.3$ bin and the $z>1.2$ bin, we get $\sigma/(1+z)$ of 0.016 and 0.053 respectively. At the lower redshift end, this is better than EAZY (see Figure\ref{fig:zaccuracy}) but in the higher redshift bin this is much worse than EAZY, as expected given the effect of blended IRAC photometry therein. While there are clearly advantages and disadvantages to both approaches, we chose our EAZY redshifts for our analysis in particular because of this behavior at $z>1$. This comparison highlights the significant biases that can exist between photometric redshifts derived based on different photometric catalogs and using different approaches. Therefore we re-emphasize that the list of overdensities we present here are only candidates. Their confirmation requires high spectroscopic completeness out $z\sim1.5$ as expected from the PFS galaxy evolution survey which will include the XMM-LSS field \citep[e.g.][]{TanakaEtAl2017}.   

\bibliography{Krefting2019_revised}

\end{document}